\documentclass[aip,rsi,amsmath,amssymb,reprint,]{revtex4-1}
\usepackage{graphicx}
\usepackage{dcolumn}
\usepackage{bm}
\usepackage{xcolor}

\begin{document}

\preprint{AIP/123-QED}

\title{Topology and complexity of the hydrogen bond network in classical models of water}

\author{Fausto Martelli}
\affiliation{IBM Research Europe, Hartree Centre, Daresbury, WA4 4AD, United Kingdom}
\email{fausto.martelli@ibm.com}
\affiliation{Department of Physics and CNR Institute of Complex Systems, Sapienza University of Rome, P.le Aldo Moro 5, 00185 Roma, Italy}
\email{fausto.martelli@ibm.com}

\begin{abstract}
Over the years, plenty of classical interaction potentials for water have been developed and tested against structural, dynamical and thermodynamic properties. On the other hands, it has been recently observed (F. Martelli et. al, \textit{ACS Nano}, \textbf{14}, 8616--8623, 2020) that the topology of the hydrogen bond network (HBN) is a very sensitive measure that should be considered when developing new interaction potentials. Here we report a thorough comparison of 11 popular non polarizable classical water models against their HBN, which is at the root of water properties. We probe the topology of the HBN using the ring statistics and we evaluate the quality of the network inspecting the percentage of broken and intact HBs. For each water model, we assess the tendency to develop hexagonal rings (that promote crystallization at low temperatures) and pentagonal rings (known to frustrate against crystallization at low temperatures). We then introduce the \emph{network complexity index}, a general descriptor to quantify how much the topology of a given network deviates from that of the ground state, namely of hexagonal or cubic ice. Remarkably, we find that the network complexity index allows us to relate, for the first time, the dynamical properties of different water models with their underlying topology of the HBN. Our study provides a benchmark against which the performances of new models should be tested against, and introduces a general way to quantify the complexity of a network which can be transferred to other materials and that links the topology of the HBN with dynamical properties. Finally, our study introduces a new perspective that can help in rationalizing the transformations among the different phases of water and of other materials.
\end{abstract}


\maketitle

\section{Introduction}\label{sec:intro}
On our planet, water is the only substance that can be found co-existing in the solid, liquid and vapour phases outside research laboratories. Its molecular simplicity hides a remarkably wide list of anomalous behaviors that stretch over the the most complex phase diagram of any pure substance~\cite{salzmann2019advances}, and whose origin lies in a critical point located at low temperatures and low pressures~\cite{mio_nature,sellberg_2014_nature,debenedetti2020,KimmelReversible2020,experimental2020}. Nonetheless, water plays a primary role in many industrial, biological and geological processes. Therefore, there is a great interest in developing classical interaction potentials able to embrace water's complex nature. This intent is aggravated by the wide span of thermodynamic conditions at which water exists (from low temperatures and low pressures of the interstellar medium to high pressures and high temperatures in the core of planets), and by the broad range of timescales required for several processes to occur (from heterogeneous nucleation to protein folding to geological processes).

After a long-lasting debate~\cite{poole_nature,liu_2010,limmer1,wikfeldt_2011,palmer_2013,limmer2,mio_nature,limmer3,chandler_rosica,martelli_comment2,martelli_comment,martelli_density}, it is now accepted that liquid water is a mixture~\cite{mio_nature,sellberg_2014_nature,martelli_comment,debenedetti2020,experimental2020,KimmelReversible2020} of molecules whose local neighborhood constantly change between an ordered tetrahedral state with local lower density, and a more distorted tetrahedral state with local higher density~\cite{martelli_connection,ShiAnomalies2020,ShiDirect2020,ShiCommon2018,RussoUnderstanding2014,RussoWater2018,santra_2015,huang_2009,nilsson_2015,wikfeldt_2011,demarzio_2017,martelli_unravelling}. These two local structures continuously interchange with each other, giving rise to a complex network of bonds that actively determines the properties of water at the macro scale~\cite{martelli_unravelling}. The percentage of such local environments depends on the thermodynamic conditions~\cite{martelli_unravelling,wikfeldt_2011}, and eventually liquid water becomes a 1:1 mixture at low temperatures and pressures~\cite{mio_nature,debenedetti2020}.

Classical interaction potentials have been developed over the years to reproduce experimental structural, dynamical and/or thermodynamic properties of water (over a relatively reduced set of thermodynamic points), and several studies have compared them against a plethora of observables~\cite{martelli_density,comparison_1,comparison_2,comparison_3,comparison_4_tip3p_tip4p,comparison_5,comparison_6,comparison_7,comparison_8,DixWhy2018}. On the other hand, even though Bernal and Fowler first recognized almost 100 years ago that water molecules build a complex network of bonds that is at the heart of water's anomalous behavior~\cite{bernal_fowler}, the pivotal role of the hydrogen bond network (HBN) has so far been showcased only in a handful of important cases involving transformations between complex phases of water~\cite{TseMechanism1999,martonak_2004,martonak_2005,mio_nature,shephard2017high,martelli_searching,martelli_unravelling}, the mutual interactions between water and biological membranes~\cite{martelli_acsnano}, and between water and graphene sheets for technological purposes~\cite{martelli_graphene}. In particular, the inspection of the topology of the HBN in biological environments is opening new avenues in enhancing the efficacy of new drugs/vaccines~\cite{martelli_redefining}. The HBN has never been used as a metric to test water models. Two reasons for this lack of comparison are: (i) it is very hard to have a direct experimental description of the HBN in liquid phases, and (ii) it is hard to probe the HBN from a computational perspective. \\

Here we fill the gap of the point (ii) above by comparing the HBN of 11 popular and widely adopted classical water models at ambient conditions. We study the TIP3P~\cite{comparison_4_tip3p_tip4p} the SPC~\cite{spc}, the SPC/E~\cite{spce} and the flexible SPC~\cite{spcflex,spcflex2} as 3-points models; the TIP4P~\cite{comparison_4_tip3p_tip4p}, the TIP4P-Ice~\cite{tip4pIce}, the TIP4P/2005~\cite{tip4p2005}, the flexible TIP4P/2005~\cite{tip4p2005flex} and the TIP4P-Ew~\cite{tip4pEw} as 4-points models; the TIP5P~\cite{tip5p} and the TIP5P-Ew~\cite{tip5pEw} as 5-points models. We probe the topology of the HBN using the ring statistics, a theoretical tool that has seen increasingly high relevance in determining the properties of bulk water~\cite{martelli_LOM,martelli_searching,martelli_unravelling,martelli_rings,mio_nature,santra_2015,leoni_2019,camisasca_proposal,martonak_2004,martonak_2005,russo_2014,mio_nature,fitzner_ice,ShiImpact2018}, of aqueous solutions~\cite{rings_solutions1,rings_solutions2,rings_solutions3,LiUnraveling2020} and of water under confinement~\cite{martelli_acsnano,martelli_graphene}. We then measure the quality of the HBN for each water model in terms of broken and intact HBs, a measure intimately linked to the fluidity and tetrahedrality of water~\cite{distasio_2014,martelli_acsnano}. \\
We also introduce the concept of \emph{network complexity index} $\xi$ that allows to quantify how much a given HBN deviates from the HBN of water in the crystalline phase at low temperatures and ambient pressure, i.e., hexagonal or cubic ice. As a showcase, we apply this index $\xi$ to one of the rings definition and counting scheme here adopted. Remarkably, the index $\xi$ allows us to link dynamical properties of water with the topology of the underlying HBN and the corresponding structural properties. Finally, we show that the HBN topology is more sensitive to the size of the simulation box with respect to, e.g., structural properties measured \emph{via} the two-bodies pair correlation function. Therefore, the inspections of the network topology should always be considered when simulating network-forming materials. This issue becomes particularly relevant when dealing with \emph{ab initio} molecular dynamics simulations which are restricted to small simulation cells. 

The article is organized as follows. In Section~\ref{sec:comput} we report the details of the numerical simulations, the ring definitions and counting schemes, and we introduce the network complexity index. In Section~\ref{sec:results} we report our main findings for all 11 water models here inspected. Conclusions and final remarks are reported in Section~\ref{sec:conclusions}.

\section{Computational details}\label{sec:comput}
In this section we describe the numerical setup, the protocols implemented to count rings and we introduce the definition of the network complexity index.

\subsection{Numerical simulations}
Our study is based on classical molecular dynamics simulations 
of systems composed of $N=1100$ water molecules described by different 11 interaction potentials in the isobaric ($N$$p$$T$) ensemble. We have employed Nos\'e-Hoover thermostat~\cite{nose,hoover} with 0.2 ps relaxation time to maintain constant temperature at $T=300$~K, and Parrinello-Rahman barostat~\cite{parrinello_rahman} with 2 ps relaxation time to maintain constant pressure at 1~bar. We have performed simulations with the GROMACS 18.0.1 package~\cite{gromacs}. All simulations have been equilibrated for 1~ns. The production runs achieved 3~ns. For each water model we have averaged over 10 independent trajectories. Our analysis investigates the properties of liquid water in the presence of thermal noise.

\subsection{Ring statistics and network complexity index}\label{sec:xi}
\begin{figure}
  \begin{center}
   \includegraphics[scale=.38]{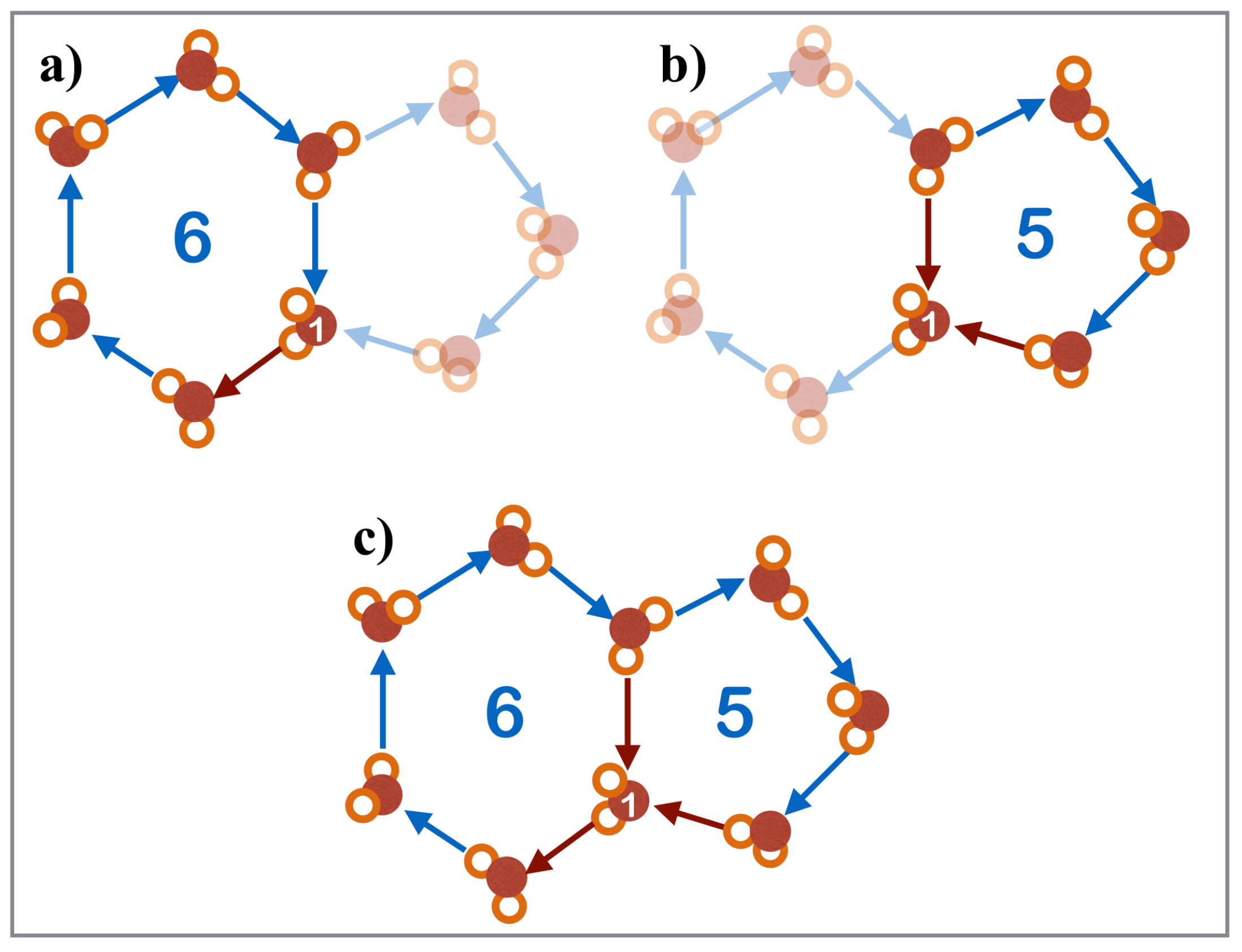}
    \caption{\label{fig:rings} Schematic representation of the three counting schemes. Red filled circles represent oxygen atoms, while red empty circles represent hydrogen atoms. The network of HBs is represented by the arrows Water molecule labeled as 1 is the starting molecule from which rings are counted. a): Molecule donates a HB (red arrow) and the shortest ring is the hexagonal one. b): Molecule 1 can accept a HB (one of the two red arrows) and the shortest ring is the pentagonal one c): Molecule 1 can either donate one HB or accept a HB, generating both the short hexagonal and pentagonal ring.}
  \end{center}
\end{figure}

In order to compute the ring statistics it is necessary to follow two steps. First of all, it is necessary to define the link between atoms/molecules. Possible definitions can be based on the formation of bonds, interaction energies, geometric distances, etc.. The second step is the definition of ring and the corresponding counting scheme. This task is of particular relevance in directional networks, like water or silica, were the donor/acceptor nature of the bonds breaks the symmetry in the linker search path. Several definitions of rings and counting schemes have been reported in the literature~\cite{king,rahman_hydrogen,guttman_ring,franzblau_computation,wooten_structure,yuan_efficient,leroux_ring}. Such different definitions have yielded to different interpretations of numerical results even for the simplest crystalline structures, not to mention more complicated networks such as amorphous silicate structures~\cite{jin_structural,hobbs_local,guttman_ring,marians_characterization,marians_local,yuan_efficient} and water~\cite{fitzner_ice,camisasca_proposal,martelli_unravelling,santra_2015}. In the case of water, these inconsistencies have been recently reconciled by Formanek and Martelli showing that they were caused by different counting schemes~\cite{martelli_rings}. \\
Here, we report a thorough analysis of the HBN for the 11 classical non-polarizable models of water. This benchmark study adopts the three ring definitions and counting schemes reported in Ref.~\cite{martelli_rings}. 
The definition of HB follows the geometric construction described in Ref.~\cite{chandler_HB}. In this regard, any quantitative measure of HBs in liquid water is somewhat ambiguous, since the notion of an HB itself is not uniquely defined. However, qualitative agreement between the definition here adopted and several other proposed definitions have been deemed satisfactory over a wide range of thermodynamic conditions, including the one here investigated~\cite{prada_2013,shi_2018_2}.
We construct rings by starting from a tagged water molecule (marked as 1 in fig.~\ref{fig:rings}) and recursively traversing the HBN until the starting point is reached again or the path exceeds the maximal ring size considered (12 water molecules in our case). In the first scheme sketched in fig.~\ref{fig:rings} a), molecule 1 \emph{donates} a HB emphasized by the red arrow, and we restrict our counting only to the shortest rings~\cite{king}, i.e., to rings that can not be further decomposed into smaller ones (the 6-folded ring in this case). We will refer to this scheme to as d1. This scheme emphasizes the directional nature of the HBs resulting in an enhanced hexagonal character of the HBN~\cite{martelli_rings}. It is therefore well suited for investigating phenomena such as nucleation, where hexagonal rings are the elemental building blocks of the final, crystal network. In the second scheme sketched in fig.~\ref{fig:rings} b), we consider only rings formed when molecule 1 \emph{accepts} a HB and we restrict our counting only to the shortest ring~\cite{king} (the 5-folded ring in this case). We will refer to this scheme to as d2. This scheme emphasizes the formation of pentagonal rings in the network~\cite{martelli_rings}, which are known to play an important role at supercooled conditions frustrating against crystallization~\cite{RussoUnderstanding2014,ShiCommon2018,martelli_unravelling}, as well as in promoting the crystallization of clathrate structures~\cite{LiClathrate2020}. These two counting schemes d1 and d2 provide drastically different distributions~\cite{martelli_rings} because of the different energies involved in accepting and donating a HB. As we will show in the forthcoming discussion, such difference translates into distinct percentages of coordination defects of the kind $\textit{A}_2\textit{D}_1$ and $\textit{A}_1\textit{D}_2$, where $\textit{A}_x\textit{D}_y$ indicates that a water molecules accepts $x$ and donates $y$ HBs. In the third scheme sketched in fig.~\ref{fig:rings} c) we loosen the previous restrictions. Molecule 1 can now either accept or donate a HB, hence counting both the 6-folded and the 5-folded ring. In the following, we will refer to this counting scheme to as d3. \\
As shown in Ref.~\cite{martelli_rings}, these three different counting schemes carry different, but complementary physical information and, therefore, allow us to make a proper comparison on the topology of the HBN generated by different classical models of water.

When studying the topology of a network via the ring statistics, the probability distribution P(n) of having a n-folded ring is a normalized quantity that does not reflect the overall number of each n-folded ring. On the other hand, the actual number of rings is important to understand the degree of complexity in a network (for a given number of molecules). The stable crystalline phase of water at ambient pressure is the hexagonal(cubic) ice, I$_{h(c)}$, characterized by an hexagonal network of bonds. In a sample of liquid water at higher temperatures, the thermal noise allows water molecules to explore a larger configurational space and, hence, the underlying HBN hosts also shorter and longer rings. Intuitively, such network is more "complex" with respect to the HBN of the ground state, and its fluctuations are related to the dynamical properties (translational and rotational diffusion) of water molecules. Knowing the symmetry of the HBN at the ground state, we can  measure the deviation from it. We introduce the \emph{network complexity index} $\xi$ defined as the ratio between the number of 6-folded rings and the total number of rings:
\begin{equation}
    \xi=\frac{n_6}{\sum_{i=3}^{12}n_i}
    \label{eq:index}
\end{equation}
where $n_i$ is the number of the $i-$th ring. The sum on the denominator of eq.~\ref{eq:index} runs from n=3, the shortest ring length possible, to n=12, the longest ring length here considered. The network complexity index $\xi$ encodes the number of rings in a given network. For the ground state I$_{h(c)}$, the network complexity index is trivially $\xi=1$. We can define the case of maximal disorder in the network when each ring length occur with the same frequency in the network. This case corresponds to a flat distribution in P(n) and, since in our case we consider 10 possible ring lengths (from n=3 to n=12), $\xi=0.1$. Here, we compute $\xi$ for the counting scheme d3, scheme c) in fig.~\ref{fig:rings} which, compared to the other schemes, accounts for the presence in the network of longer rings. 

The definition of network complexity index can be extended to other networks with different ground states (i.e., non-hexagonal networks), in which case the nominator in eq.~\ref{eq:index} will be $n_j$, $j$ being the length of the characteristic ring length at the ground state. However, when comparing the index $\xi$ in different simulations, care must be taken in order to ensure that (i) the same definition of ring and counting schemes are used, (ii) the same maximum search path (or maximal ring size) is implemented. 

It is worthy to remark, at this point, that our analysis occurs in the presence of thermal noise. As recently shown by Montes de Oca et al., the percentage of broken and intact bonds drastically changes upon removal of the thermal noise~\cite{MontesStructural2020}, i.e., in correspondence with the inherent potential energy surfaces (IPES). While we do expect quantitative differences in correspondence with the IPES with respect to the results that we will show in Section~\ref{sec:results}, we are confident that the overall trend should be preserved.

\section{Results}\label{sec:results}
In this section, we report and discuss our results in terms of topology and complexity of the HBNs, as well as their quality in terms of broken and intact HBs for the water 11 models. We then compare the network complexity indices and we relate them to the dynamical properties of each water model. Finally, we show and discuss the effects of the simulation box size on the topology of the HBN.

\subsection{3 points models}
\begin{figure}
  \begin{center}
   \includegraphics[scale=.33]{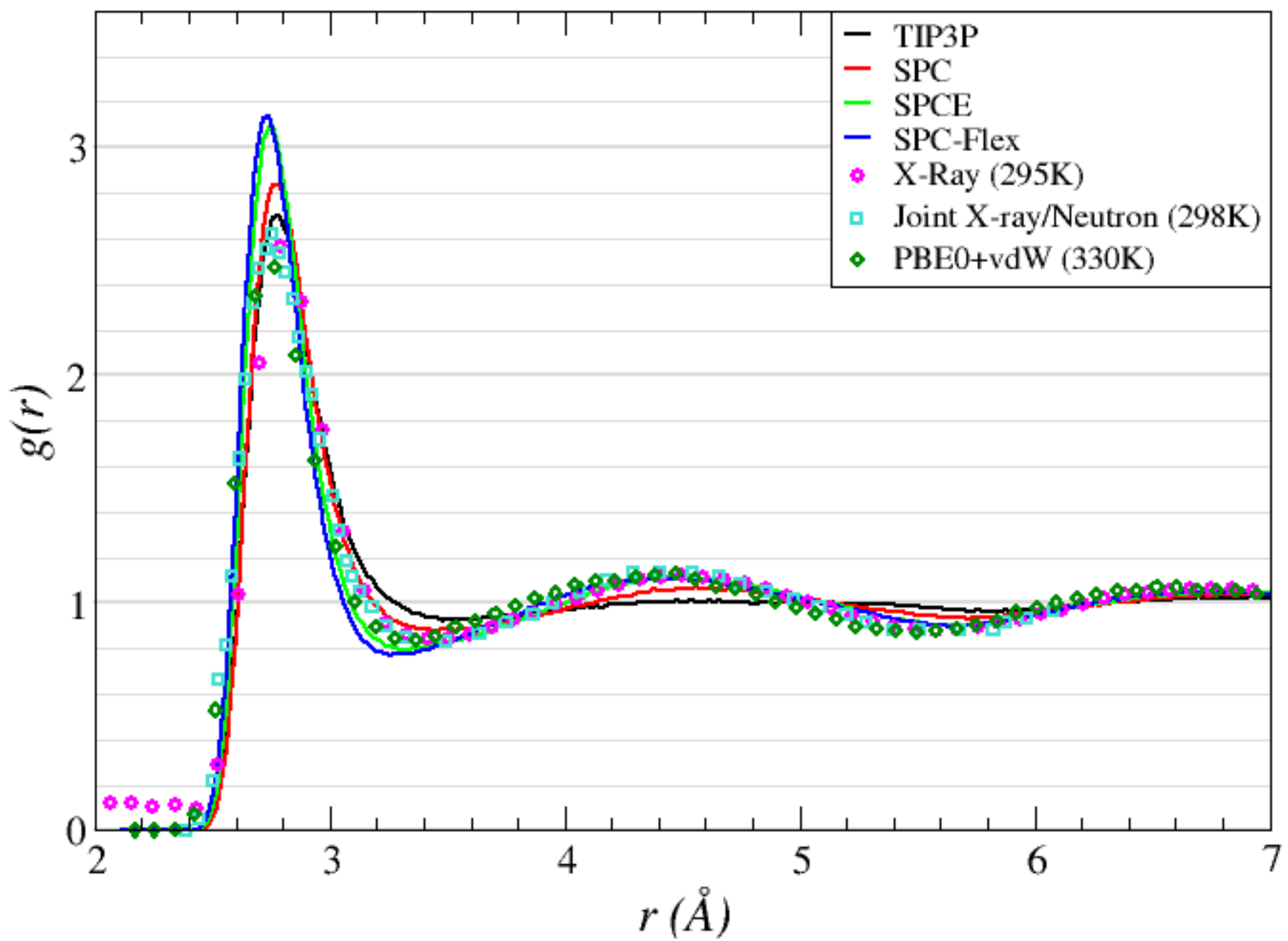}
    \caption{\label{fig:3pointsRDF} The oxygen-oxygen two-bodies pair correlation, g$_2$(r), of liquid water for the TIP3P (black), SPC (red), SPC/E (green) and the flexible SPC model (blue). The g$_2$(r) obtained from various scattering experiments~\cite{skinnerExp,soperExp} and \emph{ab initio} molecular dynamics simulations~\cite{distasio_2014} are reported for comparison with open symbols.}
  \end{center}
\end{figure}

we start our analysis by comparing, in fig.~\ref{fig:3pointsRDF}, the oxygen-oxygen two-bodies pair correlation functions g$_2$(r) for 3-points models with the g$_2$(r) obtained from various scattering experiments~\cite{skinnerExp,soperExp} (open circles and squares) and \emph{ab initio} molecular dynamics (AIMD) simulations~\cite{distasio_2014} (open triangles) at the PBE0 level of theory accounting for non-local van der Waals/dispersion interactions~\cite{distasio_2014}. It is worth to mention, at this point, that the PBE0+vdw level of theory gives an accurate g$_2$(r) compared to the experimental one, but does not capture the correct density difference between liquid water at ambient conditions and I$_h$. \\
We can observe that the TIP3P model (black line) has the first peak at $\sim2.8$~nm with intensity comparable with that of scattering experiments and \emph{ab initio} simulations. On the other hand, after the first minimum the distribution is almost flat with no peaks at larger distances. This is indicative of a lack of structurization at larger distances. The SPC model (red line) is able to perform better than the TIP3 model, with a g$_2$(r) showing hints of a second and a third peak, though shifted compared to the experimental and \emph{ab initio} g$_2$(r). On the other hand, the intensity of the first peak overcomes the experimental and AIMD first peak. The SPC/E model (green line) results in a better g$_2$(r) in terms of intensity and position of the second and third peak with respect to the experimental and the \emph{ab initio} g$_2$(r), reflecting the better performances in density and diffusion constant than the SPC model~\cite{spce}. On the other hand, the intensity of the first peak is further enhanced. The flexible SPC water model (blue line) is a re-parametrization of the SPC water model in which the O--H stretching is made anharmonic, and thus the dynamical behavior is well described and bulk density and permettivity are correct~\cite{praprotnik2004}. The g$_2$(r) of the flexible SPC model is characterized by a high intensity first peak and a deeper first minimum with respect to the rigid SPC model, and correctly captures the profile of the g$_2$(r) at larger distances.

Overall, the sequence TIP3, SPC, SPC/E and flexible SPC is characterized by an increment in the height of the first peak that also shifts slightly towards lower distances, a corresponding deepening of the first minimum and a gradual appearance of a second and third peak that tend to overlap with the experimental and with AIMD ones for the SPC/E and the flexible SPC models.

\begin{figure}
  \begin{center}
   \includegraphics[scale=.33]{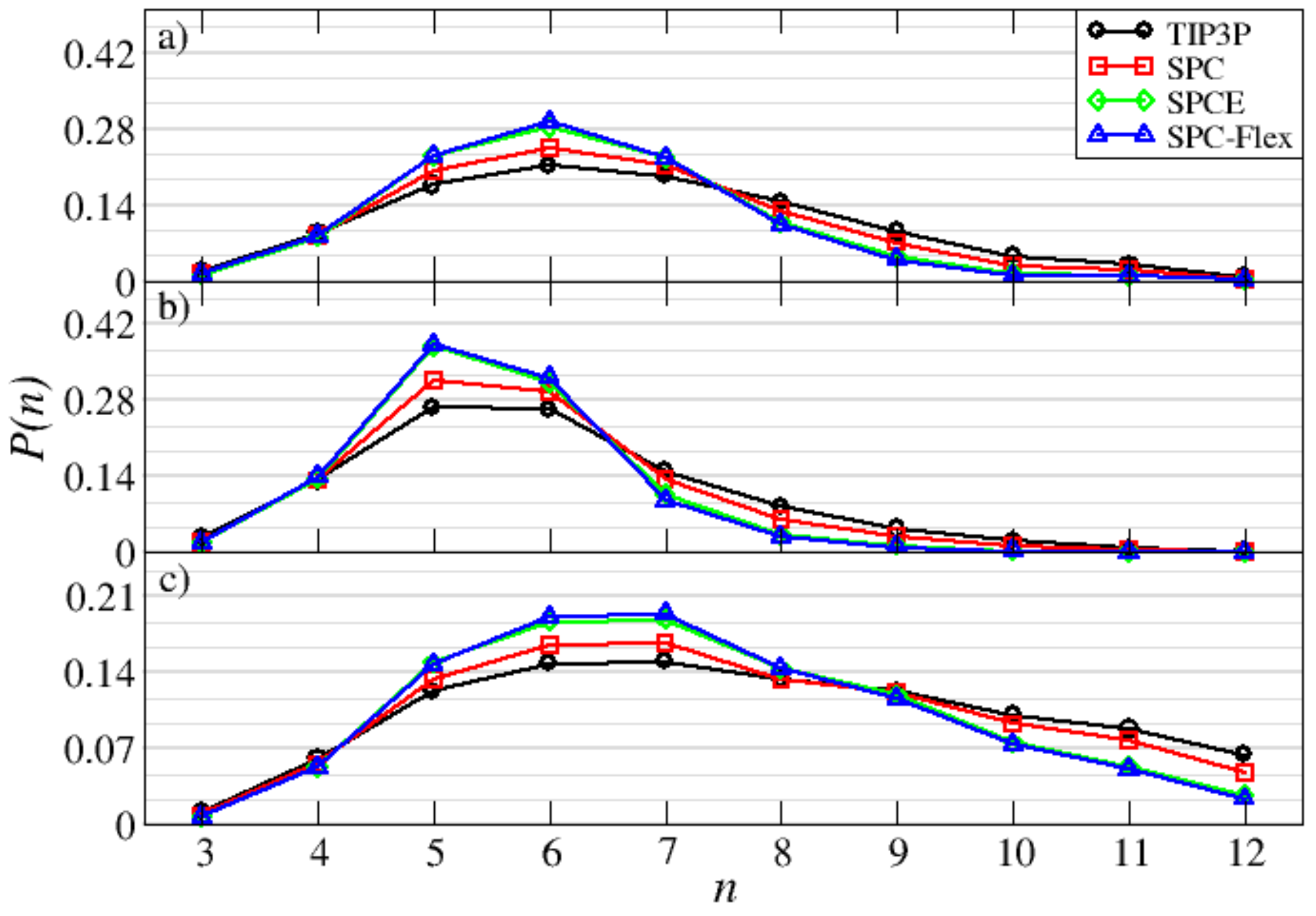}
    \caption{\label{fig:3pointsRings} Probability distributions of the hydrogen-bonded n-folded rings, P(n), for liquid water at ambient conditions described by the TIP3P (black open circles), the SPC (red open squares), the SPC/E (green open diamonds) and the flexible SPC (blue open triangles) models. Panel a) reports the P(n) according to the definition sketched in fig.~\ref{fig:rings} a); panel b) reports the p(n) according to the definition sketched in fig.~\ref{fig:rings} b); panel c) reports the P(n) according to the definition sketched in fig.~\ref{fig:rings} c).}
  \end{center}
\end{figure}

In fig.~\ref{fig:3pointsRings} we report the probability distribution P(n) of having a n-folded ring, with n$\in[3,12]$ for the classical 3 points models. The distribution for the TIP3P model is reported as open circles, SPC as open squares, SPC/E as open diamonds and flexible SPC as open triangles. It is worth to remark, at this point, that the P(n) is a normalized distribution and, therefore, it does not reflect the actual number of each ring. \\
Panel a) reports the P(n) according to the definition d1 sketched in fig.~\ref{fig:rings} a). According to this counting scheme that emphasizes the directionality of the HBs, all networks have a dominating hexagonal character. Such character is milder in the TIP3P model (black open circles), which shows a very broad P(n) and whose network accommodates also 10- and 11-folded rings. The hexagonal character of the network grows moving to the SPC model (red open squares), with a corresponding reduction of longer (n$>$8) rings. The reduction of longer rings develops along with an enhancement of 5- and 7-folded rings, which are comparable to 6-folded rings in terms of energy~\cite{camisasca_proposal}. In particular, the HBN of the SPC model described with this counting scheme is almost completely deprived of rings with n$>$10. The hexagonal character of the network further grows in the SPC/E (green open diamonds) and in the flexible SPC (open blue triangles). These models show almost identical P(n)s. In particular, besides the enhanced hexagonal character, the HBNs of these models have an enhanced pentagonal and heptagonal character. Consequently, the contribution from longer rings, namely rings with n=8 and n=9, is less marked. Therefore, loosening the holonomic constraints and allowing the (re-parametrized) SPC model to vibrate has a major effect not just on the structural properties (as shown from the g$_2$(r), fig.~\ref{fig:3pointsRDF}) but also on the topology of the HBN.\\
Panel b) reports the P(n) according to the definition d2 sketched in fig.~\ref{fig:rings} b). As shown in Ref.~\cite{martelli_rings}, this counting scheme emphasizes the pentagonal character of the HBN, as the starting water molecule must accept one HB instead of donating. Interestingly, the TIP3P model shows, according to this counting scheme, an almost equal pentagonal and hexagonal character with a tail populating configurations up to n=10. The HBN of the SPC model, on the other hand, shows an improved pentagonal character with a slight increase in the hexagonal character and a reduction of longer rings which mostly disappear at n$>$9. The pentagonal character of the HBN over the hexagonal one become particularly dominant in the SPC/E and in the flexible SPC models, for which rings characterized by n$>$8 are mostly absent. \\
Panel c) reports the P(n) according to the definition d3 sketched in fig.~\ref{fig:rings} c). Since this counting scheme does not discern among the donor/acceptor character of the HBs and does not implement the shortest path criterion~\cite{king}, the resulting topology is more complex with respect to the previous counting schemes. Therefore, for this scheme we also compute the network topology index $\xi$ (eq.~\ref{eq:index}) and report them in table~\ref{tab:table1}. According to this counting scheme, the HBN for all the inspected models show a similar character in terms of hexagonal and heptagonal rings. The TIP3P model shows a quite broad distribution with a considerable contribution of longer (n$>$7) rings. The presence of rings with n=12 reflects the more complex topology of the network with respect to the previous counting schemes. Interestingly, such distribution is comparable with that of the TIP3P model optimized~\cite{tip3pcharmm} for biological simulations~\cite{martelli_acsnano}. Because of the very broad character of P(n), the TIP3P model has a low value of network complexity index $\xi=0.1496$, not too far from the value $\xi=0.1$ that characterizes (as described in Section~\ref{sec:xi}) a flat distribution P(n) with equal populations. Moving to the SPC model, we observe a slight enhancement of n=5, and a more pronounced enhancement of n=6 and n=7 which, as for the TIP3P model, equally characterize the HBN. Contrarily, the network is deprived of longer rings, namely of n=10, n=11 and n=12 rings. The slight increase in n=6 causes a small increment in the network complexity index to $\xi=0.1665$. Moving to the SPC/E model, we observe a mild increment in n=5, with a relevant increment in n=6 and n=7, as well as a marked depletion of longer rings. The complexity index for the SPC/E model further increases to $\xi=0.1863$. A similar distribution characterizes the HBN of the flexible SPC model, which almost overlaps with the P(n) of the SPC/E model. The distribution of the flexible SPC model suggests that, as mentioned above, the introduction of flexibility plays a very important role in shaping the network. The complexity index for the flexible SPC model rises to $\xi=0.1922$. 
\begin{table}
  \begin{center}
    \label{tab:table1}
    \begin{tabular}{c|c|c|c|c}
       & TIP3P & SPC & SPC/E & SPC-Flex \\  
       \hline
       $\xi$ & 0.1496 & 0.1665 & 0.1863 & 0.1922 \\
      \hline
    \end{tabular}
    \caption{Values of the network complexity index $\xi$ computed for the rings counting scheme d3 for the 3 points models of water.}
  \end{center}
\end{table}

\begin{figure}
  \begin{center}
   \includegraphics[scale=.33]{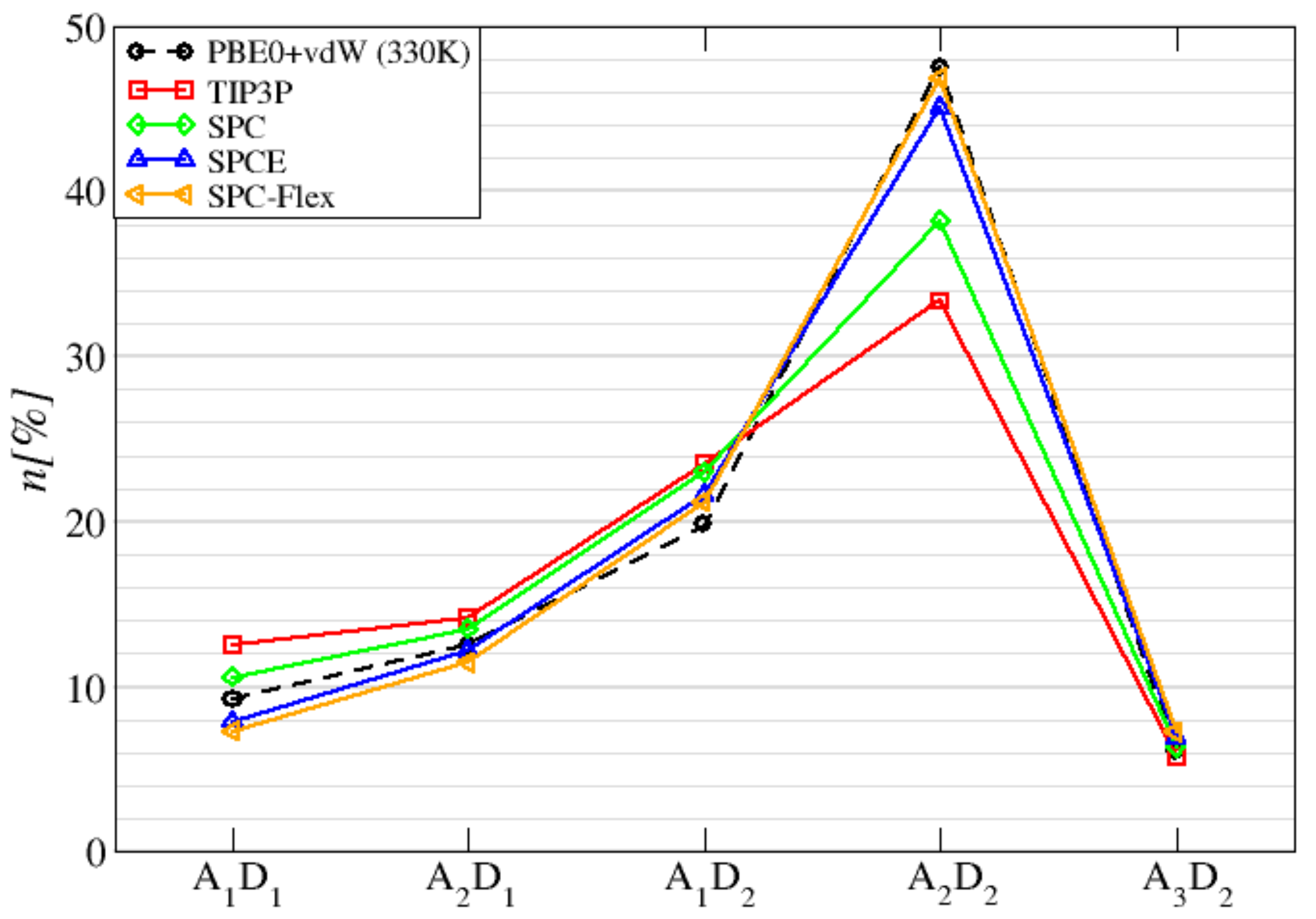}
    \caption{\label{fig:3points} Percentage-wise decomposition of the intact HBs per water molecule into acceptor-(A) and donor-(D) for \emph{ab initio} liquid water at T=330~K as black open circles, and for the 3 points models. The TIP3P model is reported as red open squares, the SPC as green open diamonds, the SPC/E as blue open triangles and the flexible SPC as orange open left triangles. The $x$-axis labels $\textit{A}_x\textit{D}_y$ indicate the number of acceptor ($\textit{A}_x$) and donor ($\textit{D}_y$) HBs. For clarity we omit combinations  with minor contributions, \textit{e.g.}, $\textit{A}_3\textit{D}_1$, $\textit{A}_0\textit{D}_y$, $\textit{A}_x\textit{D}_0$, \textit{etc.} }
  \end{center}
\end{figure}

In fig.~\ref{fig:3points}, we report the percentage of broken and intact HBs for the 3 point models of water and, for comparison, the \emph{ab initio} liquid water at the PBE0 level of theory with vdW long range interactions at a temperature of 330K to account for nuclear quantum effects~\cite{distasio_2014}. We adopt the following syntax: $\textit{A}_x\textit{D}_y$ indicates the number of acceptors ($\textit{A}_x$) and donors ($\textit{D}_y$) HBs. The network in \emph{ab initio} liquid water (black open circles) is dominated by a intact HBs ($\textit{A}_2\textit{D}_2$), which account for $\sim48\%$. The second highest configuration is the $\textit{A}_1\textit{D}_2$, with $\sim20\%$, followed by $\textit{A}_2\textit{D}_1$ ($12\%$), $\textit{A}_2\textit{D}_2$ ($\sim10\%$) and $\textit{A}_3\textit{D}_2$ ($\sim5\%$). The percentage of this last configuration, i.e., the configuration $\textit{A}_3\textit{D}_2$, is shared with all 3 points models. The percentage of broken and intact HBs in the TIP3P model (red open squares) quantitatively differs from the distribution of \emph{ab initio} liquid water, but qualitatively shows a similar behaviour. In particular, the distribution for the TIP3P model is dominated by a markedly reduced amount of intact HBs with a percentage $\textit{A}_2\textit{D}_2\sim33\%$, followed by $\textit{A}_1\textit{D}_2$ with $\sim23\%$, $\textit{A}_2\textit{D}_1$ with $\sim12\%$ and an almost equal percentage of $\textit{A}_1\textit{D}_1$. The low percentage of intact HBs explains the very broad distribution of rings (fig.~\ref{fig:3pointsRings}) as well as the absence of structurization beyond the first hydration shell in the g$_2$(r) (fig.~\ref{fig:3pointsRDF}). Moving from the TIP3 model to the SPC model (open green diamonds) we observe a slight decrease in the configuration $\textit{A}_1\textit{D}_1$ to $\sim10\%$ and an increment in the percentage of intact HBs to $\sim37\%$. Therefore, the SPC model is characterized by a slightly enhanced percentage of fully coordinated configurations which cause an enhancement in 5-, 6-, and 7-folded rings, as reported in fig.~\ref{fig:3pointsRings}, and are responsible for the appearance of the second peak in the g$_2$(r) (fig.~\ref{fig:3pointsRDF}). The modified SPC models, namely the SPC/E (open blue triangles) and the flexible SPC model (open orange left triangles), are both characterized by a distribution that mostly overlap with the \emph{ab initio} liquid water potential. The high percentage of intact HBs reaches values in the order of $\sim45-48\%$ for both SPC/E and the flexible SPC models, and this enhanced tetra-coordinated character explains the enhanced number of 5-, 6-, and 7-folded rings and the corresponding decrease in longer rings, as reported in fig.~\ref{fig:3pointsRings}, as well as an enhancement in the intensity of the first peak in the g$_2$(r) and a corresponding deepening of the following minimum.

\subsection{4 points models}
\begin{figure}
  \begin{center}
   \includegraphics[scale=.33]{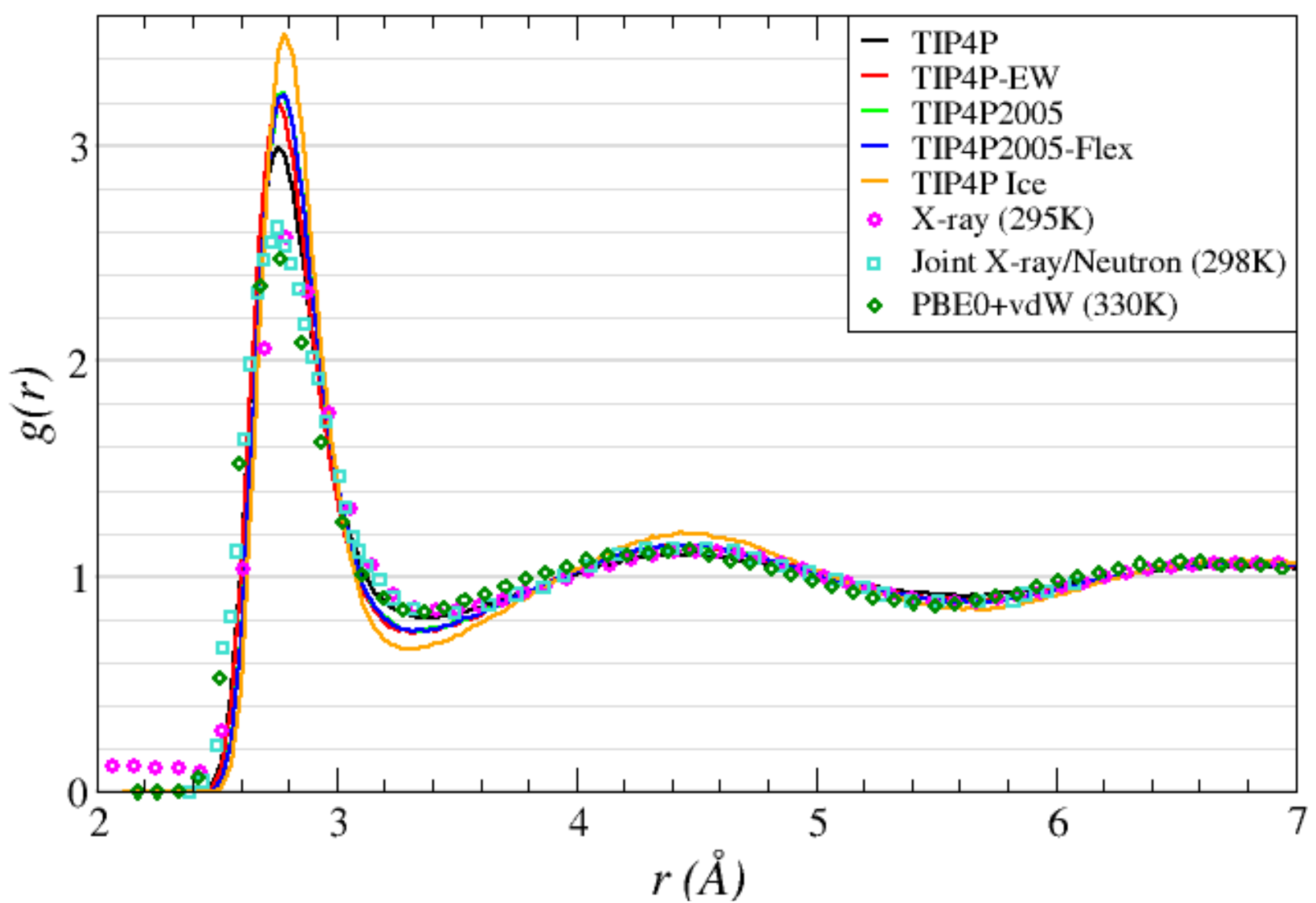}
    \caption{\label{fig:4pointsRDF} The oxygen-oxygen two-bodies pair correlation, g$_2$(r), of liquid water for the TIP4P (black), TIP4P-Ew (red), TIP4P/2005 (green), the flexible TIP4P/2005 (blue), and the TIP4P-Ice (orange) models. The g$_2$(r) obtained from various scattering experiments~\cite{skinnerExp,soperExp} and \emph{ab initio} molecular dynamics simulations~\cite{distasio_2014} are reported for comparison with open symbols.}
  \end{center}
\end{figure}

In fig.~\ref{fig:4pointsRDF} we report the g$_2$(r) for the 4 points models, namely the TIP4 (black line), the TIP4P-Ew (red line), the TIP4P/2005 (green line), the flexible TIP4P/2005 (blue line) and the TIP4P-ice (orange line) models. As for the 3 points models, the open symbols represent the g$_2$(r) from experimental data~\cite{skinnerExp,soperExp} and from \emph{ab initio} molecular dynamics simulations~\cite{distasio_2014}, and serve as a comparison. The TIP4P model is characterized by the intensity of the first peak as high as $\sim3$, the closest to the experimental/\emph{ab initio} ones. The depth of the following minimum is slightly more pronounced than the experimental/AIMD. At larger distances, on the other hand, the g$_2$(r) almost overlap with the experimental/AIMD ones. The TIP4P-Ew (red line) has been optimized for simulating water in biological environments~\cite{tip4pEw}. With respect to the TIP4P model, the TIP4P-Ew shows a higher intensity in the first peak of the g$_2$(r), reaching $\sim3.2$. Other features of the g$_2$(r) are qualitatively overlapping with the experimental/AIMD distribution functions. A similar trend characterizes the g$_2$(r) for the TIP4P/2005 (green line) and for the flexible TIP4P/2005 (blue line) which are mostly indistinguishable from the distribution function of the TIP4P-Ew function. The TIP4P-Ice, on the other hand, shows a g$_2$(r) more structured with respect to the previous ones. The intensity of the first peak reaches values as close as $\sim3.5$, and the following minimum shows a deeper depth. Finally, the intensity of the second peak is slightly more pronounced with respect to the experimental/AIMD ones. Such "over"-structurization is not surprising considering that this water model has been developed to study crystalline and non crystalline solid forms of ice~\cite{tip4pIce}.

Overall, moving from the TIP4P model to the TIP4P-Ew, TIP4P/2005, flexible TIP4P/2005 and the TIP4P-Ice we observe a systematic enhancement of the first peak which also slightly shifts towards larger values. All models well capture the second and third peaks.

\begin{figure}
  \begin{center}
   \includegraphics[scale=.33]{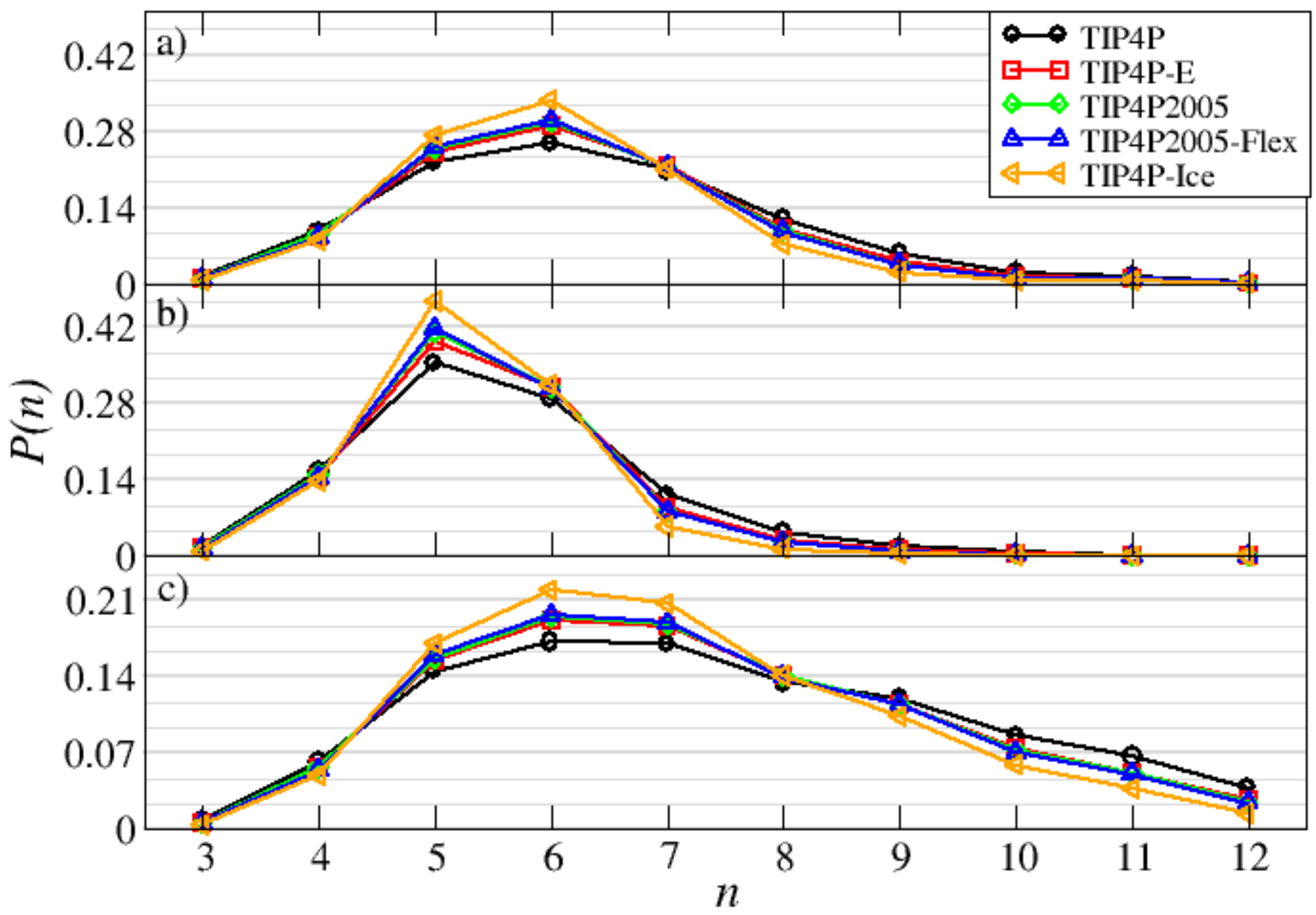}
    \caption{\label{fig:4pointsRings} Probability distributions of the hydrogen-bonded n-folded rings, P(n), for liquid water at ambient conditions described by the TIP4P (black open circles), the TIP4P-EW (red open squares), the TIP4P/2005 (green open diamonds), the flexible TIP4P/2005 (blue open triangles), and the TIP4P-Ice (open orange triangles) models.}
  \end{center}
\end{figure}

In fig.~\ref{fig:4pointsRings} we report the probability distribution P(n) of having a n-folded ring, with n$\in[3,12]$ for the TIP4P (black open circles), the TIP4P-Ew (red open squares), the TIP4P/2005 (green open diamonds), the flexible TIP4P/2005 (blue open triangles) and the TIP4P-Ice (orange open left triangles) models. In panel a) we report the P(n) according to the definition d1 sketched in fig.~\ref{fig:rings} a). All classical models of water show a P(n) maximized at n=6. In particular, the TIP4P model shows, among all the 4 points models here inspected, the broader distribution, with an almost equal contribution of 5-folded and 7-folded rings. Interestingly, the distribution for the TIP4P model is less broad compared with the distribution for the TIP3P model reported in fig.~\ref{fig:3pointsRings} a), with almost no rings longer than n=9. The HBN of the TIP4P-Ew model shows a slight increase in the pentagonal and in the hexagonal character and a corresponding lower character in rings longer than n=7. Very similar distributions occur for the TIP4P/2005 and for the flexible TIP4P/2005 models. The hexagonal character of the HBN is further emphasized in the TIP4P-Ice model for which we can observe also a slight increment in the pentagonal character and a depletion of rings longer than n=7. This enhanced hexagonal character of the HBN is a consequence of the parametrization of this model which has been optimized to reproduce crystalline and non-crystalline solid forms of ice.\\
In panel b) of fig.~\ref{fig:4pointsRings} we report the P(n) computed using the rings and counting scheme definition d2 sketched in fig.~\ref{fig:rings} b). According to such counting scheme, all distributions are maximized over n=5. In particular, the P(n) for the TIP4P model shows the broader distribution, with some contribution from n=8, while the contributions to the HBN from longer rings are mostly negligible. The TIP4P-Ew model shows an enhanced pentagonal character and a slightly increment of the hexagonal character as well, with a reduction of the contribution coming from rings longer than n=6. Such tendency becomes more pronounced moving to the TIP4P/2005 and to the flexible TIP4P/2005 models, while differences among these three models are minimal. On the other hand, the pentagonal character of the HBN is particularly enhanced in the TIP4P-Ice model whose HBN, according to this counting scheme, dose not account for rings longer than n=7.\\ 
In panel c) we report the P(n) computed using the counting scheme d3 sketched in fig.~\ref{fig:rings} c). For this scheme, we also compute the complexity indices $\xi$, reported in table~\ref{tab:table2}. We can observe that the distribution P(n) for the TIP4P model is fairly broad and mostly dominated by an almost equal amount of 6- and 7-folded rings followed by 5-folded rings, while the amount of longer rings decreases with increasing the rings lengths. The complexity index for the TIP4P model is $\xi=0.1721$, which is larger compared to the TIP3 and the SPC models, but lower compared to the other 3 points models here inspected. The network of TIP4P-Ew model is characterized by a consistent increment in the hexagonal character, followed by the heptagonal character and a smaller increment in the pentagonal character. Longer rings weight less with respect to the TIP4P model. The complexity index for the TIP4P-Ew is $\xi=0.1887$, larger compared to the TIP4P model. The distribution P(n) for the TIP4P/2005 and the flexible TIP4P/2005 models mostly overlap with the P(n) for the TIP4P-Ew, and have a similar complexity index, namely $\xi=0.1915$ for the TIP4P/2005 model and $\xi=0.1984$ for the flexible TIP4P/2005 model. Interestingly, the P(n) for the TIP4P-Ew, for the TIP4P/2005 and for the flexible TIP4P/2005 models show a breaking in the equal hexagonal and heptagonal character of the network, with a slight predominance of the hexagonal character which justifies the higher values of the index $\xi$. The network of the TIP4P-Ice model is characterized by a further enhancement of the hexagonal character, followed by the heptagonal and hexagonal characters, with a depletion of longer rings. Such enhancement in the hexagonal character is reflected by the complexity index $\xi=0.2155$, the closest to the value of ice among all models inspected in this work. 

The similar distributions of rings for the TIP4P-Ew, TIP4P/2005 and flexible TIP4P/2005 in all counting schemes suggest that such models perform equally in terms of HBN, and this reflects the very similar g$_2$(r) (fig.~\ref{fig:4pointsRDF}). On the other hand, the TIP4P-Ice, which shows a more structured g$_2$(r) (fig.~\ref{fig:4pointsRDF}), tends to favour a network with a shorter connectivity.
\begin{table}
  \begin{center}
    \label{tab:table2}
    \begin{tabular}{c|c|c|c|c|c}
       & TIP4P & TIP4P-Ew & TIP4P/2005 & TIP4P/2005-Flex & TIP4P-Ice \\  
       \hline
 $\xi$ & 0.1721 & 0.1887 & 0.1915 & 0.1984 & 0.2155 \\
      \hline
    \end{tabular}
    \caption{Values of the network complexity index $\xi$ computed for the rings counting scheme d3 for the 4 points models of water.}
  \end{center}
\end{table}

\begin{figure}
  \begin{center}
   \includegraphics[scale=.33]{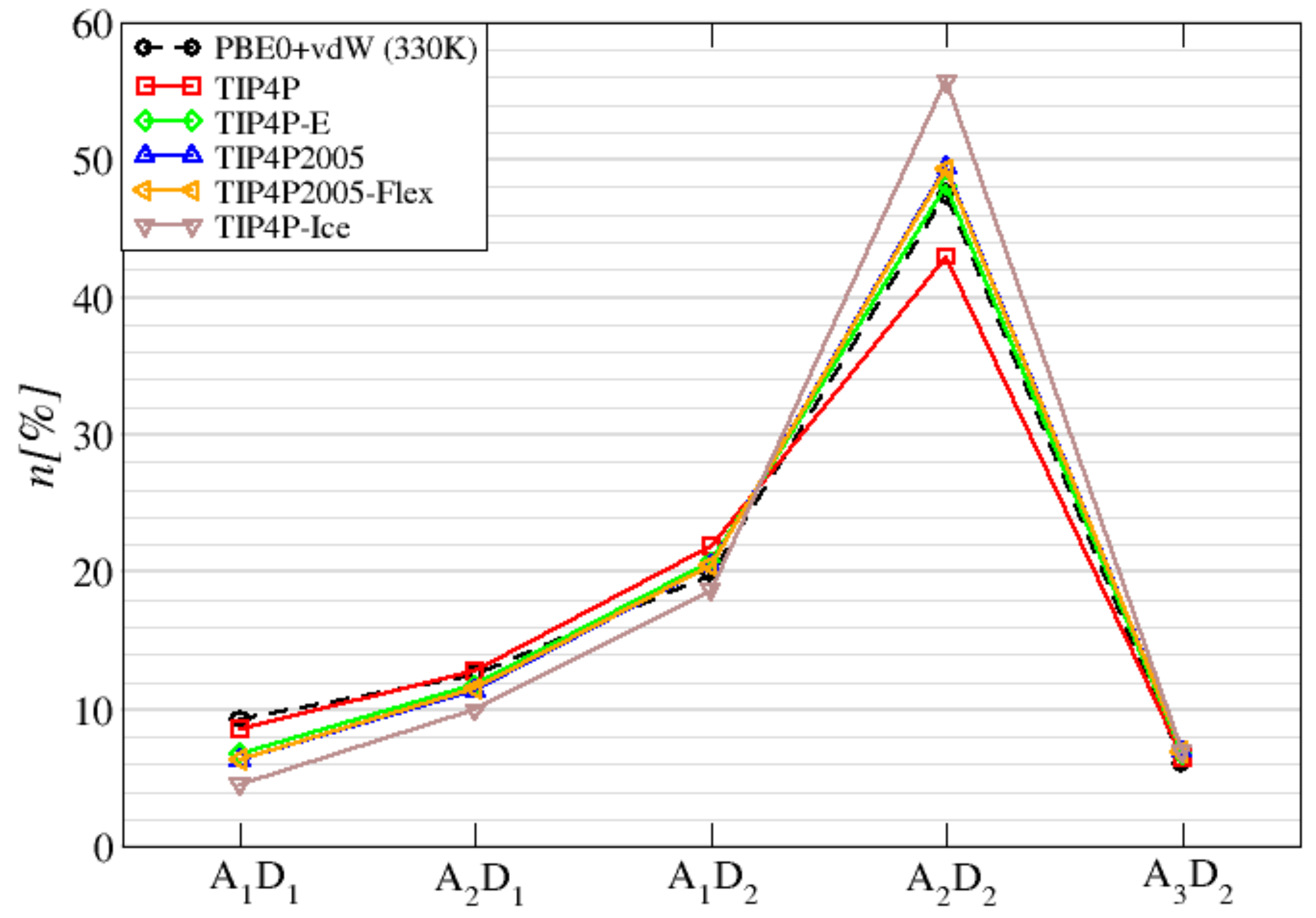}
    \caption{\label{fig:4points} Percentage-wise decomposition of the intact HBs per water molecule into acceptor-(A) and donor-(D) for \emph{ab initio} liquid water at T=330~K as black open circles, and for the 4 points models. The TIP4P model is reported as red open squares, the TIP4P-Ew model as green open diamonds, the TIP4P/2005 model as blue open triangles, the flexible TIP4P/2005 as orange open left triangles, and the TIP4P-Ice model as brown open lower triangles. The distribution for the TIP4P/2005 model almost perfectly overlaps with the distribution for the flexible TIP4P/2005 model.}
  \end{center}
\end{figure}

In fig.~\ref{fig:4points} we report the percentage of broken and intact HBs for the 4 points classical models and for \emph{ab initio} liquid water. We can observe that the percentage of broken and intact HBs for all 4 points models qualitatively resembles that of \emph{ab initio} liquid water (open black circles). With respect to AIMD water, the TIP4P model (open red squares) underestimates the percentage of $\textit{A}_2\textit{D}_2$ configurations to $\sim43\%$ and a slightly higher percentage of $\textit{A}_1\textit{D}_2$ configurations ($\sim22\%$). All other configurations mostly overlap with the configurations of AIMD water. The percentage of broken and intact HBs for the TIP4P-Ew (open green diamonds), for the TIP4P/2005 (open blue triangles) and the flexible TIP4P/2005 (open orange left triangles) mostly overlap with each other, reflecting the almost overlapping distributions of rings and the similar values of the complexity index $\xi$ (fig.~\ref{fig:4pointsRings}). In particular, we observe almost no differences between the distribution for the TIP4P/2005 and the flexible TIP4P/2005 models. With respect to the TIP4P model, they recover the percentage of $\textit{A}_2\textit{D}_2$ for the AIMD liquid water, while showing a small reduction of $\textit{A}_1\textit{D}_1$ defects ($\sim7\%$). Such increment in the percentage of $\textit{A}_2\textit{D}_2$ to $\sim48\%$ reflects the stronger 5-, 6- and 7-folded character of the HBN described in fig.~\ref{fig:4pointsRings}. The TIP4P-Ice model, on the other hand, overestimate the percentage of intact HBs with respect to AIMD water, reaching $\sim55\%$ of the total configurations. This enhanced 4-folded coordination explains the distribution of rings (fig.\ref{fig:4pointsRings}) describing an HBN particularly enriched in 6- and 7-folded rings, as well as the higher intensity in the first peak of the g$_2$(r) (fig~\ref{fig:4pointsRDF}).

Overall, we can state that, overall, the 4 points models are characterized by similar HBNs, with the exception of the TIP4P model which shows a broad distribution of rings caused by a lower percentage of intact HBs, and the TIP4P-Ice model, whose high percentage of intact HBs causes generates a network characterized by a strong hexagonal and heptagonal character with a marked reduction of longer rings.

\subsection{5 points models}
\begin{figure}
  \begin{center}
   \includegraphics[scale=.33]{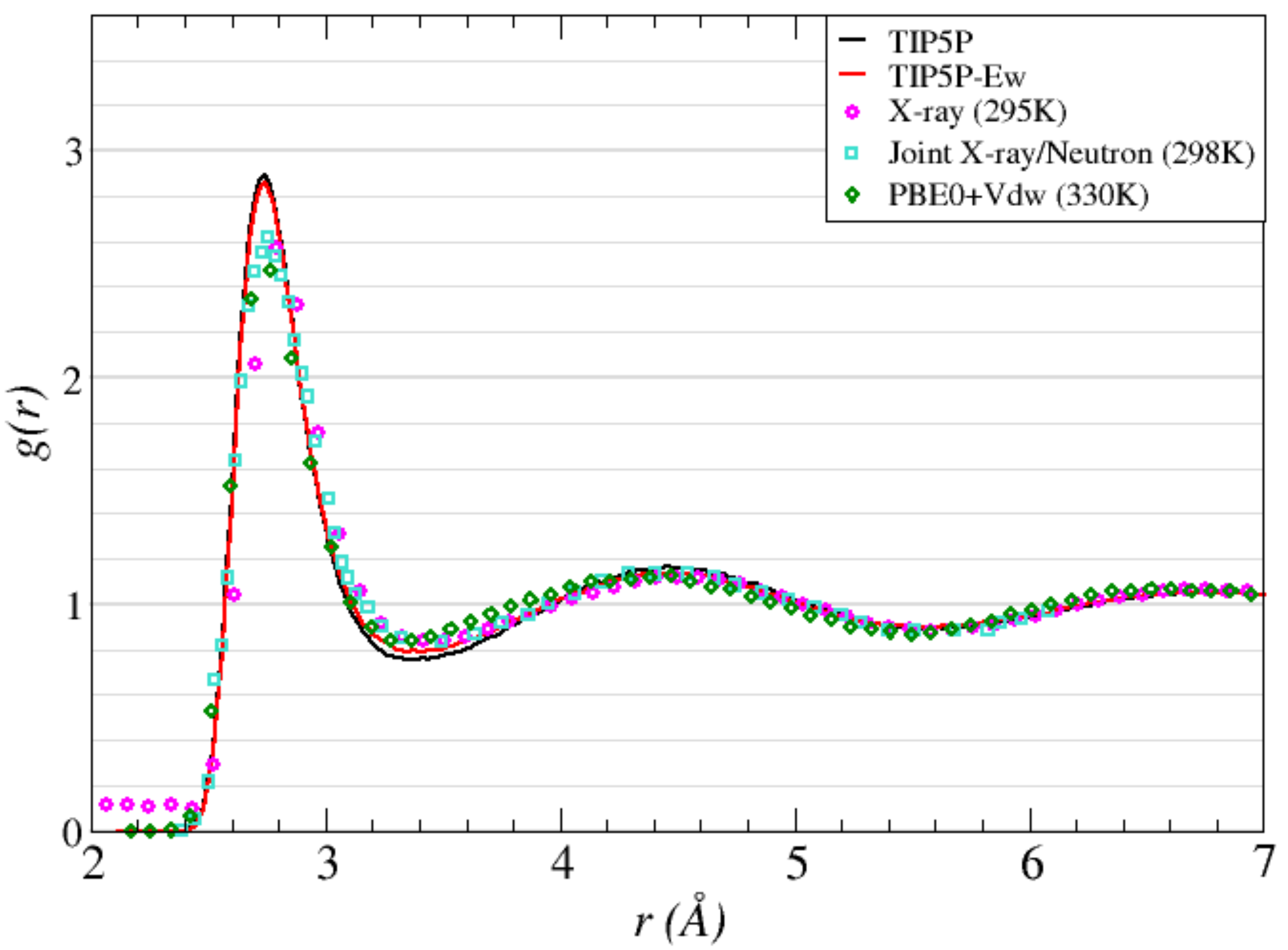}
    \caption{\label{fig:5pointsRDF} The oxygen-oxygen two-bodies pair correlation, g$_2$(r), of liquid water for the TIP5P (black) and the TIP5P-EW (red) models. The g$_2$(r) obtained from various scattering experiments~\cite{skinnerExp,soperExp} and \emph{ab initio} molecular dynamics simulations~\cite{distasio_2014} are reported for comparison with open symbols.}
  \end{center}
\end{figure}

In fig.~\ref{fig:5pointsRDF} we compare the g$_2$(r) of two 5 points models, namely the TIP5P (black line) and the TIP5P-E (red line) models with the g$_2$(r) obtained from various scattering experiments~\cite{skinnerExp,soperExp} and \emph{ab initio} molecular dynamics simulations~\cite{distasio_2014} reported as open symbols. We can observe that the g$_2$(r) for both 5 points models mostly overlap. With respect to the 3 points and to the 4 points models, the g$_2$(r) of both 5 points models are more closer to the experimental and to the AIMD g$_2$(r). The intensity of the first peak for the TIP5P and the TIP5P-E models is below 3, namely $\sim$2.9. The depth of the first minimum is slightly more pronounced with respect to the experimental and the AIMD g$_2$(r), while experimental and AIMD peaks at longer distances are well captured.

\begin{figure}
  \begin{center}
   \includegraphics[scale=.33]{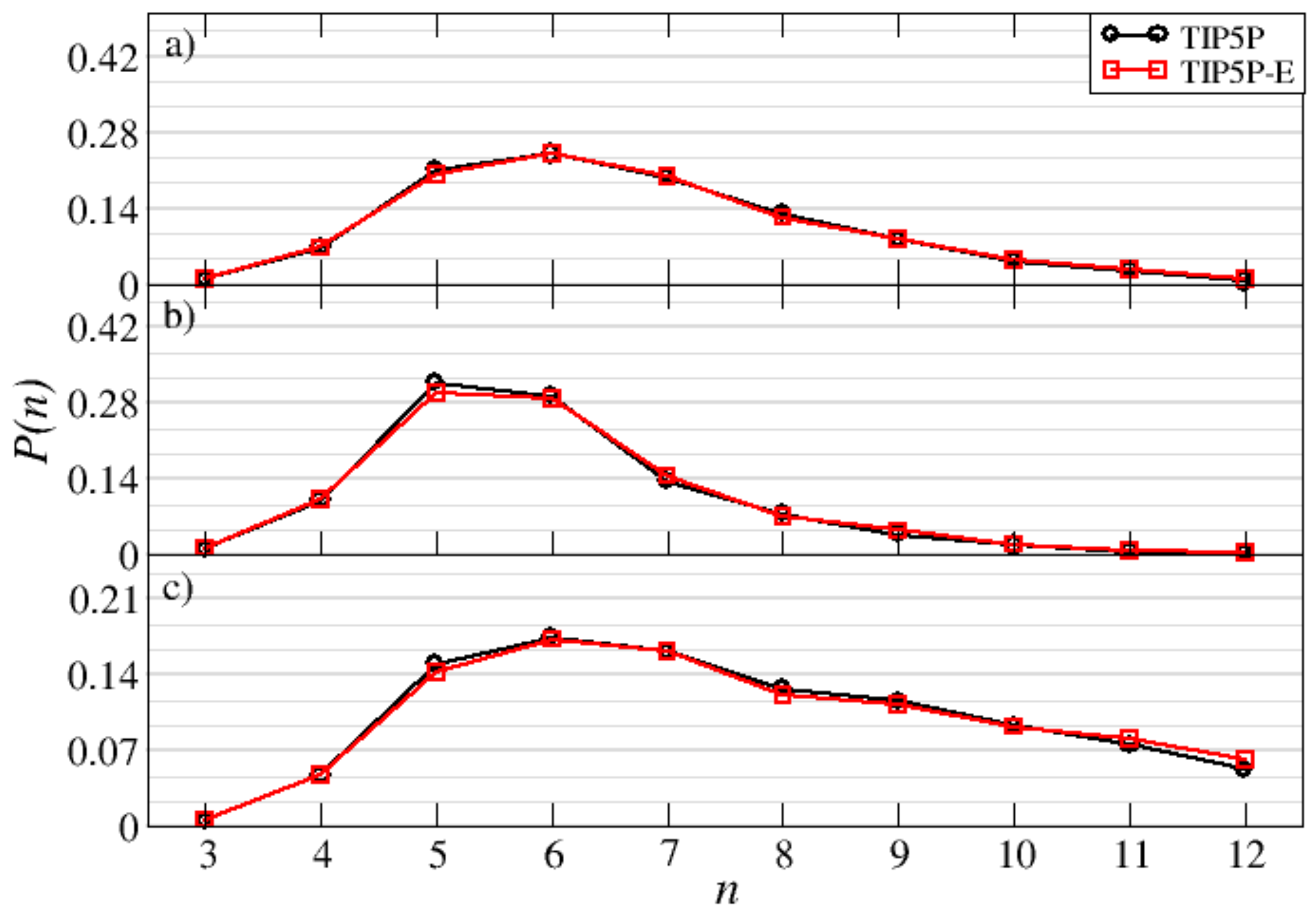}
    \caption{\label{fig:5pointsRings} Probability distributions of the hydrogen-bonded n-folded rings, P(n), for liquid water at ambient conditions described by the TIP5P (black open circles) and the TIP5P-EW (red open squares).}
  \end{center}
\end{figure}

In fig.~\ref{fig:5pointsRings} we report the probability distribution P(n) of having a n-folded ring, with n$\in[3,12]$ for the TIP5P model (open black circles) and for the TIP5P-E model (open red squares). In panel a) we report the P(n) computed using the counting scheme d1 sketched in fig.~\ref{fig:rings} a) and which emphasizes the directionality of the HBN. Both the TIP5P and the TIP5P-E models provide similar distributions which are slightly maximized at n=6, and contributions of rings up to n=11.\\
In panel b) we report the P(n) computed using, as a counting scheme, the definition d2 sketched in fig.~\ref{fig:rings} b). As for the previous case, also in this case the two distributions are mostly indistinguishable and with an almost equal pentagonal and hexagonal character. Overall, the pentagonal character of both HBNs is less pronounced with respect to the four points models, and qualitatively resemble the P(n) for the three points model TIP3P (fig.~\ref{fig:3pointsRings} b)).\\
In panel c) we report the P(n) computed using the definition d3 and sketched in fig.~\ref{fig:rings} c). For this counting scheme, we also compute the complexity indices reported in table~\ref{tab:table3}.
As for the previous counting schemes, the differences between the two P(n)'s are minimal, indicating that the two model provide similar HBNs. Such similarity can be quantified observing that both models have almost the same complexity index $\xi$, i.e., $\xi=0.1774$ for the TIP5P model and $\xi=0.1754$ for the TIP4P-Ew model. The values of $\xi$ for both networks are comparable with that of the TIP4P model but, with respect to the TIP4P model, the hexagonal character of the HBN is more pronounced for both the TIP5P and the TIP5P-E models. Besides the TIP4P-Ice model, among all other interaction potentials here inspected, the TIP5P and the TIP5P-E models are the only one for which the counting scheme d3 is (slightly) maximized towards n=6.
\begin{table}
  \begin{center}
    \label{tab:table3}
    \begin{tabular}{c|c|c}
       & TIP5P & TIP5P-E  \\  
       \hline
 $\xi$ & 0.1774 & 0.1754 \\
      \hline
    \end{tabular}
    \caption{Values of the network complexity index $\xi$ computed for the rings counting scheme d3 for the 5 points models of water.}
  \end{center}
\end{table}

\begin{figure}
  \begin{center}
   \includegraphics[scale=.33]{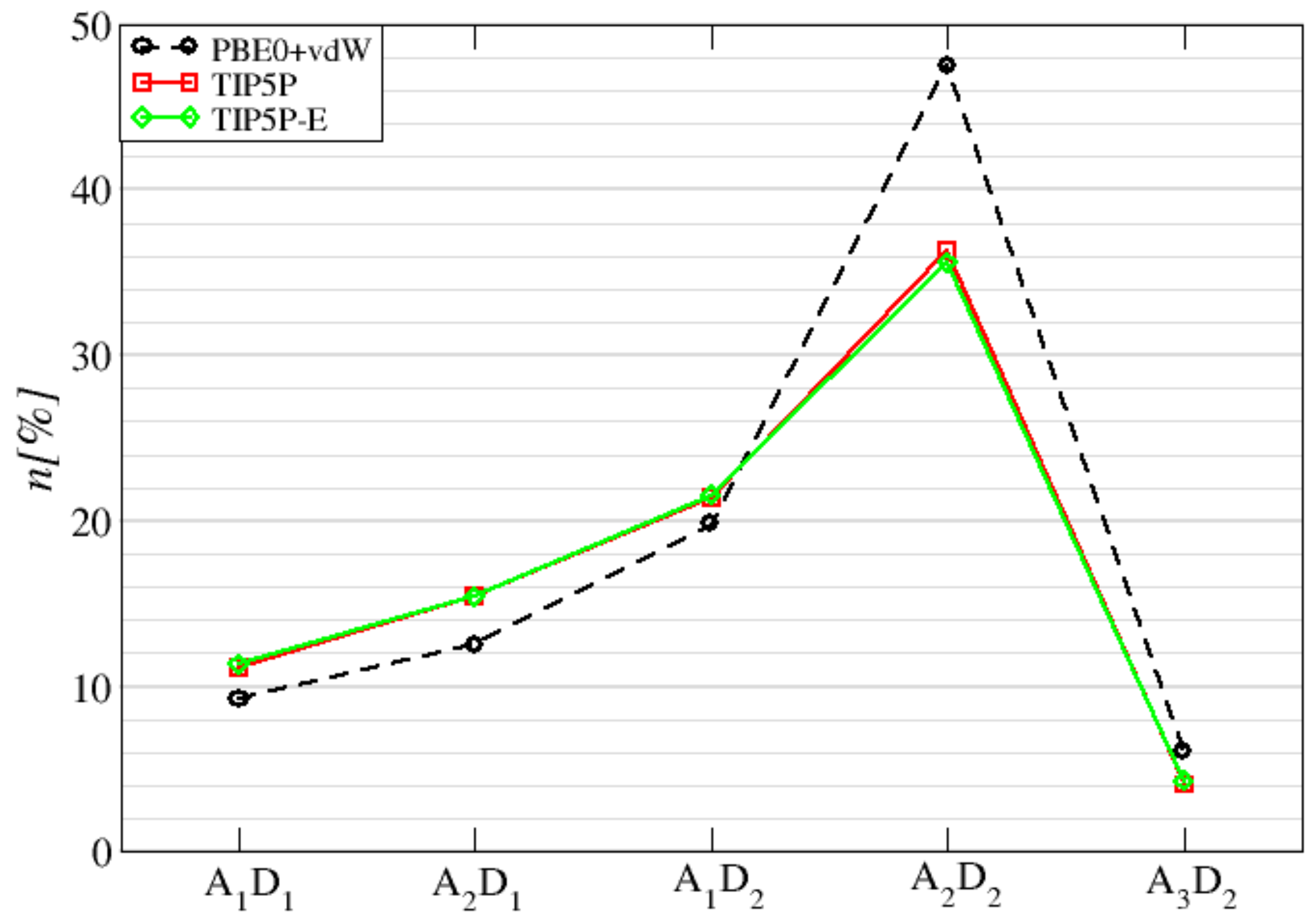}
    \caption{\label{fig:5points} Percentage-wise decomposition of the intact HBs per water molecule into acceptor-(A) and donor-(D) for \emph{ab initio} liquid water at T=330~K as black open circles, and for the 5 points models. The TIP5P model is reported as red open squares and the TIP5P-EW model as green open diamonds.}
  \end{center}
\end{figure}

In fig.~\ref{fig:5points} we report the percentage of broken and intact HBs for the 5 points classical models and for \emph{ab initio} liquid water. With respect to the AIMD liquid water (open black circles), both the TIP5P (open red squares) and the TIP5P-E (open green diamonds) models are characterized by a markedly lower percentage of intact HBs ($\sim36\%$), while low-coordinated defects occur in higher percentages. The $\textit{A}_1\textit{D}_2$ configuration account for the $\sim22\%$ of the total configurations, followed by $\textit{A}_2\textit{D}_1$ configurations with a percentage of $\sim16\%$ and $\textit{A}_1\textit{D}_1$ configurations with $\sim12\%$ of the total configurations. The low amount of intact HBs (compared with the AIMD water), reflects the broad distribution of rings and the corresponding low value of the network complexity indices $\xi$ (fig.~\ref{fig:5pointsRings}).
\subsection{Relation between network complexity and dynamical properties}
\begin{figure}
  \begin{center}
   \includegraphics[scale=.27]{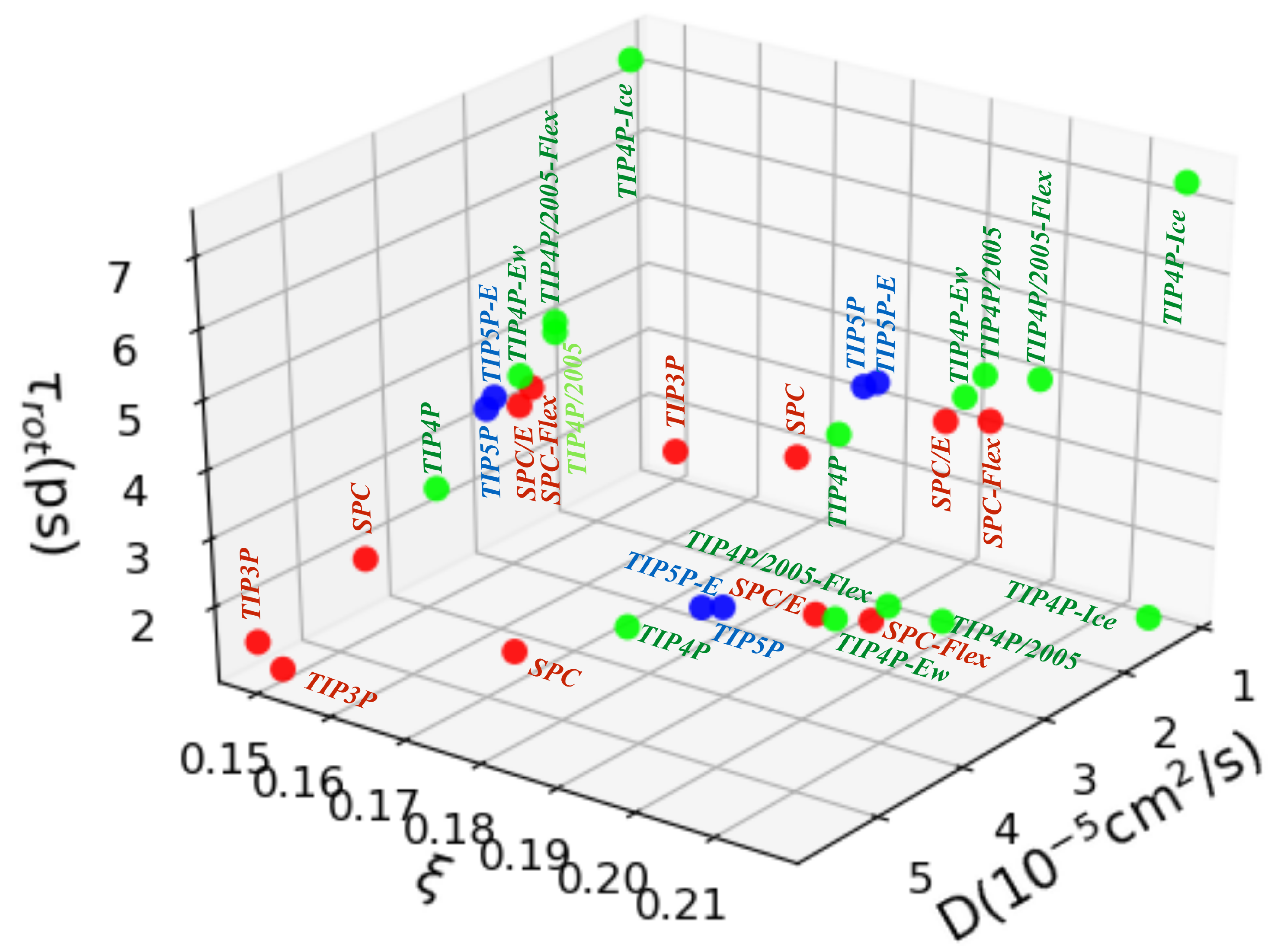}
    \caption{\label{fig:proj} Three dimensional plot reporting the projection on the corresponding two dimensions of the values acquired by the 11 classical models of water. We report the values for the network complexity index $\xi$, the diffusion coefficient and the rotational relaxation time $\tau_{rot}$. Data for 3-points models are reported in red, while data for the 4-points models are reported in green and data for the 5-points models are reported in blue.}
  \end{center}
\end{figure}

We here show how the network complexity is linked to dynamical properties, namely the translational and rotational diffusion. This link comes from the observation that bonding can be viewed as a competition between the energy gained from the formation of a bond, and the entropy loss due to the reduction in configurational volume that occurs when two particles are constrained to stay close relative to each other. The establishment of an extended network of bonds occurs when the energy gain (that controls the lifetime of bonds) balances the entropy loss. \\ 
For each water model we have computed the diffusion coefficient and the rotational relaxation time ($\tau_{rot}$), and we have reported them against the network complexity index $\xi$. We have computed the diffusion coefficient from the mean squared displacement, and the rotational relaxation time $\tau_{rot}$ from the integral of the rotational autocorrelation function as reported in Refs~\cite{CaleroStructural2016,martelli_redefining} $C_{rot}(t)=\left< \textit{OH}(t)\cdot\textit{OH}(0) \right>$, i.e., $\tau_{rot}=\int_{0}^{+\infty}C_{rot}(t)dt$. Fig.~\ref{fig:proj} show the values of these three observables in a three dimensional plot, with projections on the corresponding two dimensional spaces. The values for 3-points models are reported in red, the values for 4-points models are reported in green, and the values of 5-points models are reported in blue. We can observe a clear correlation between the complexity of the HBN and the dynamical properties of water molecules. The models with the highest diffusion coefficients and the fastest rotational relaxation times are the TIP3P and the SPC models, which are also characterized by the lowest values of the index $\xi$. Contrarily, the TIP4P-Ice model is the model with the lowest diffusion coefficient and the slowest rotational relaxation time, and the highest index $\xi$. Overall, we can observe that for all models of water the higher the value of $\xi$ (reported in ascending order in the tables~\ref{tab:table1},~\ref{tab:table2},~\ref{tab:table3}), the slower the diffusion coefficient and the rotational relaxation time. Therefore, we can assert that there is a clear correlation between the complexity of the HBN and dynamical properties, a relation never observed before. Such correlation suggests that faster diffusion and rotations allow water molecules to increase the possible connections between each other, hence increasing the configurational space that the network can explore resulting in a more complex topology able to host a larger amount of longer rings.
\subsection{System size dependence}
\begin{figure}
  \begin{center}
   \includegraphics[scale=.33]{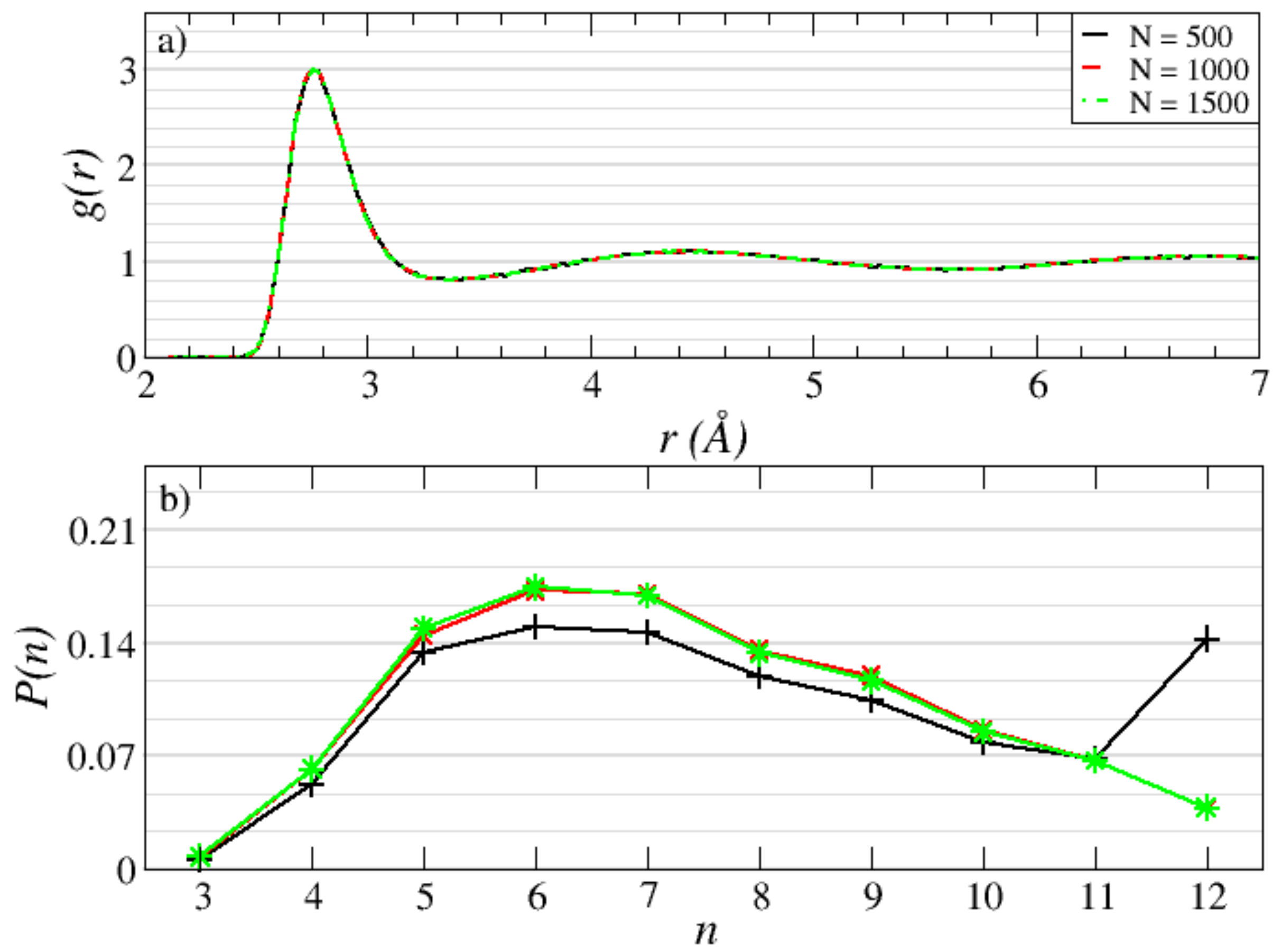}
    \caption{\label{fig:size} Panel a): system size dependence on the oxygen-oxygen two bodies pair correlation function for the TIP4P model using a box containing 500 water molecules (black continuous line), 1000 water molecules (red dashed line) and 1500 water molecules (green dotted-dashed line). Panel b): ring distribution for a system described by the TIP4P model in a box containing 500 water molecules (black pluses), 1000 water molecules (red crosses) and 1500 water molecules (green stars).}
  \end{center}
\end{figure}

We now turn our attention to the study of finite size effects, commonly inspected when computing physical quantities to check whether a system suffers from periodicity artifacts. In fig.~\ref{fig:size} a) we report the oxygen-oxygen g$_2$(r) for a simulation box containing 500 water molecules (black continuous line), 1000 water molecules (red dashed line) and 1500 water molecules (green dotted-dashed line) interacting with the TIP4P classical interaction potential. We can observe that the three g$_2$(r) perfectly overlap. In fig.~\ref{fig:size} b) we report the probability distribution P(n) of having a n-folded ring for the three cases inspected above and computed according to the ring definition and counting scheme d3. We can observe that the distribution computed for the smaller simulation box with N=500 molecules is remarkably different from the distributions with N=1000 and N=1500 molecules. In particular, we observe a strong enhancement of n=12 rings which causes a reduced contribution of shorter rings to the P(n). This result indicates that a search path of n=12 water molecules is too long for a small simulation box with only N=500 water molecules, and the increment in n=12 is caused by periodicity in the simulation box. It is worth to mention that the definitions d1 and d2 do not show such behavior (data not reported), as the network investigated with these definitions does not host rings as long as n=12 (see fig.~\ref{fig:4pointsRings} upper and middle panels). Therefore, although the g$_2$(r) for N=500 is the same as the g$_2$(r) for larger simulation boxes, care must be taken when inspecting the network topology and in how such inspection is performed.

\section{Conclusions}\label{sec:conclusions}
In this article we have tested 11 popular non polarizable classical interaction potentials for water against their hydrogen bond networks (HBNs). We have probed the topology of the HBN using three schemes that emphasize different physical features. We have evaluated the quality of the HBNs in terms of broken and intact HBs, and we have linked our results to structural properties measured \emph{via} the two bodies pair correlation function g$_2$(r). We have then introduced the network complexity index $\xi$ that measures how much the topology of a HBN deviates from that of the ground state, and we have tested it to one of the three rings counting schemes. We have shown that the index $\xi$ is directly related to dynamical properties, hence establishing a clear cause-effect relationship between molecular motions and network connectivity. Finally, we have inspected how periodicity artifacts can influence the topology of the HBN. We have performed all studies at ambient conditions, i.e., T=300~K and p=1~bar. Although different water models have (very) different melting points, their network topology --and hence their network complexity-- remain roughly unchanged away from the limit of supercooling~\cite{martelli_rings}. On the other hand, when water is under confinement, water molecules in the proximity of the surfaces undergo a drastic change in the dynamics~\cite{samatas_2018,martelli_acsnano,martelli_graphene,gallo_2010,CamisascaEffect2020,IorioGlassy2019,IorioSlow2020,TenuzzoProtein2020,CaleroWater2020} and in network topology~\cite{martelli_acsnano,martelli_graphene}. Therefore, the choice of a given interaction potential becomes of particular relevance.\\
In the class of 3 points models, we have tested the TIP3P, the SPC, the SPC/E and the flexible SPC models. We have found that the TIP3P model is characterized by the less structured network, with broad distributions or rings and a network rich in coordination defects which allows the network to arrange in long rings. In particular, the counting scheme d3 gives a low value of complexity index $\xi$ which reflects the large distance from the HBN of crystalline ice for which $\xi=1.0$. The broad rings distribution and the low percentage of intact HBs explain the absence of a second hydration peak in the g$_2$(r). The percentage of intact HBs increases in the SPC model which is, therefore, characterized by an HBN with fewer longer rings and by a g$_2$(r) with signatures of a second hydration peak. The SPC/E and the flexible SPC models are characterized by a further increase in the percentage of intact HBs, comparable with that of \emph{ab initio} liquid water. The resulting HBNs accommodate an even lower percentage of longer rings and, therefore, the network complexity index is higher compared to the previous models, indicating a closer (but still very far) HBN to the HBN of I$_{h(c)}$, in agreement with the more structured g$_2$(r). It is of particular interest to observe how the introduction of flexibility in the SPC model drastically affects the topology of the HBN.\\
In the class of 4 points models, we have tested the TIP4P, the TIP4P-Ew, the TIP4P/2005, the flexible TIP4P/2005 and the TIP4P-Ice models. The TIP4P model is the only model whose network accommodates a lower percentage of intact HBs with respect to \emph{ab initio} liquid water. The topology of the corresponding HBN is hence the most complex, i.e., with a low complexity index. The TIP4P-Ew, the TIP4P/2005 and the flexible TIP4P/2005 models are characterized by a similar percentage of intact HBs, comparable with that of \emph{ab initio} liquid water. The corresponding HBNs have comparable topologies and values of complexity indices. The TIP4P-Ice model, finally, shows a higher percentage of intact HBs with respect to \emph{ab initio} liquid water. The topology of the corresponding HBN is the less complex among all models here studied, with small contributions of longer rings and the highest value of $\xi$ index. Such results explain the over structured g$_2$(r) with respect to both \emph{ab initio} water and experimental results.\\
In the class of 5 points models, we have tested the TIP5P and the TIP5P-E models. Both models have similar HBNs, characterized by a lower percentage of broken HBs with respect to \emph{ab initio} liquid water. Both networks are fairly complex, accommodating longer rings causing low values of the network complexity index. Overall, the balance between intact HBs and network topology allows the 5 points models to be the better models in reproducing the \emph{ab initio} and experimental water g$_2$(r).

Overall, we have shown that water models endowed with the fastest dynamics are able to establish more complex networks, while models with the slowest dynamics establishes networks more closely related to that of cubic or hexagonal ice. In particular, among the 11 models here inspected the TIP3P and the SPC models are the ones with the fastest dynamics and with the network deviating the most from that of ice. On the other hand, the TIP4P-Ice model is the one with the slowest dynamics and, hence, with a network of HB the closer to that of ice.

Finally, we have shown that the topology of the HBN might be affected by finite size effects when other observables such as, e.g., the two body pair correlation function, do not show such sensitivity.

In conclusion, the topology of the HBN and its quality in terms of broken and intact HBs are more sensitive quantities than other physical observables~\cite{martelli_acsnano,martelli_graphene}. Therefore, the properties of the HBN should be inspected along with all other properties such as, e.g., structural, dynamical and thermodynamic properties when developing new interaction potentials. Our study provides a benchmark evaluated following three different ring definitions and counting schemes. New interaction potentials should be tested against such results. Nonetheless, the network complexity index provides a direct quantitative measure of how much a HBN is complex and far from the HBN at the ground state, and a direct link to the dynamical properties. Such quantity can be transferred to other materials. The effect of polarization on the topology of the HBN and on its quality should be investigated.

\begin{acknowledgments}
We acknowledge support from the STFC Hartree Centre's Innovation Return on Research programme, funded by the Department for Business, Energy and Industrial Strategy.
\end{acknowledgments} 

\linespread{0.1}
\bibliography{main}

\begin{thebibliography}{100}%
\makeatletter
\providecommand \@ifxundefined [1]{%
 \@ifx{#1\undefined}
}%
\providecommand \@ifnum [1]{%
 \ifnum #1\expandafter \@firstoftwo
 \else \expandafter \@secondoftwo
 \fi
}%
\providecommand \@ifx [1]{%
 \ifx #1\expandafter \@firstoftwo
 \else \expandafter \@secondoftwo
 \fi
}%
\providecommand \natexlab [1]{#1}%
\providecommand \enquote  [1]{``#1''}%
\providecommand \bibnamefont  [1]{#1}%
\providecommand \bibfnamefont [1]{#1}%
\providecommand \citenamefont [1]{#1}%
\providecommand \href@noop [0]{\@secondoftwo}%
\providecommand \href [0]{\begingroup \@sanitize@url \@href}%
\providecommand \@href[1]{\@@startlink{#1}\@@href}%
\providecommand \@@href[1]{\endgroup#1\@@endlink}%
\providecommand \@sanitize@url [0]{\catcode `\\12\catcode `\$12\catcode
  `\&12\catcode `\#12\catcode `\^12\catcode `\_12\catcode `\%12\relax}%
\providecommand \@@startlink[1]{}%
\providecommand \@@endlink[0]{}%
\providecommand \url  [0]{\begingroup\@sanitize@url \@url }%
\providecommand \@url [1]{\endgroup\@href {#1}{\urlprefix }}%
\providecommand \urlprefix  [0]{URL }%
\providecommand \Eprint [0]{\href }%
\providecommand \doibase [0]{http://dx.doi.org/}%
\providecommand \selectlanguage [0]{\@gobble}%
\providecommand \bibinfo  [0]{\@secondoftwo}%
\providecommand \bibfield  [0]{\@secondoftwo}%
\providecommand \translation [1]{[#1]}%
\providecommand \BibitemOpen [0]{}%
\providecommand \bibitemStop [0]{}%
\providecommand \bibitemNoStop [0]{.\EOS\space}%
\providecommand \EOS [0]{\spacefactor3000\relax}%
\providecommand \BibitemShut  [1]{\csname bibitem#1\endcsname}%
\let\auto@bib@innerbib\@empty
\bibitem [{\citenamefont {Salzmann}(2019)}]{salzmann2019advances}%
  \BibitemOpen
  \bibfield  {author} {\bibinfo {author} {\bibfnamefont {C.~G.}\ \bibnamefont
  {Salzmann}},\ }\href@noop {} {\bibfield  {journal} {\bibinfo  {journal} {J.
  Chem. Phys.}\ }\textbf {\bibinfo {volume} {150}},\ \bibinfo {pages} {060901}
  (\bibinfo {year} {2019})}\BibitemShut {NoStop}%
\bibitem [{\citenamefont {Palmer}\ \emph {et~al.}(2014)\citenamefont {Palmer},
  \citenamefont {Martelli}, \citenamefont {Liu}, \citenamefont {Car},
  \citenamefont {Panagiotopoulos},\ and\ \citenamefont
  {Debenedetti}}]{mio_nature}%
  \BibitemOpen
  \bibfield  {author} {\bibinfo {author} {\bibfnamefont {J.~C.}\ \bibnamefont
  {Palmer}}, \bibinfo {author} {\bibfnamefont {F.}~\bibnamefont {Martelli}},
  \bibinfo {author} {\bibfnamefont {Y.}~\bibnamefont {Liu}}, \bibinfo {author}
  {\bibfnamefont {R.}~\bibnamefont {Car}}, \bibinfo {author} {\bibfnamefont
  {A.~Z.}\ \bibnamefont {Panagiotopoulos}}, \ and\ \bibinfo {author}
  {\bibfnamefont {P.~G.}\ \bibnamefont {Debenedetti}},\ }\href@noop {}
  {\bibfield  {journal} {\bibinfo  {journal} {Nature}\ }\textbf {\bibinfo
  {volume} {510}},\ \bibinfo {pages} {385} (\bibinfo {year}
  {2014})}\BibitemShut {NoStop}%
\bibitem [{\citenamefont {Sellberg}\ \emph {et~al.}(2014)\citenamefont
  {Sellberg}, \citenamefont {Huang}, \citenamefont {McQueen}, \citenamefont
  {Loh}, \citenamefont {Laksmono}, \citenamefont {Sclesinger}, \citenamefont
  {Sierra}, \citenamefont {Nordlund}, \citenamefont {Hampton}, \citenamefont
  {Starodub}, \citenamefont {DePonte}, \citenamefont {Beye}, \citenamefont
  {Chen}, \citenamefont {Martin}, \citenamefont {Barty}, \citenamefont
  {Wikfeldt}, \citenamefont {Weiss}, \citenamefont {Caronna}, \citenamefont
  {Feldkamp}, \citenamefont {Skinner}, \citenamefont {Seibert}, \citenamefont
  {Messerschmidt}, \citenamefont {Williams}, \citenamefont {Boutet},
  \citenamefont {Pettersson}, \citenamefont {Bogan},\ and\ \citenamefont
  {Nilsson}}]{sellberg_2014_nature}%
  \BibitemOpen
  \bibfield  {author} {\bibinfo {author} {\bibfnamefont {J.~A.}\ \bibnamefont
  {Sellberg}}, \bibinfo {author} {\bibfnamefont {C.}~\bibnamefont {Huang}},
  \bibinfo {author} {\bibfnamefont {T.~A.}\ \bibnamefont {McQueen}}, \bibinfo
  {author} {\bibfnamefont {N.~D.}\ \bibnamefont {Loh}}, \bibinfo {author}
  {\bibfnamefont {H.}~\bibnamefont {Laksmono}}, \bibinfo {author}
  {\bibfnamefont {D.}~\bibnamefont {Sclesinger}}, \bibinfo {author}
  {\bibfnamefont {R.~G.}\ \bibnamefont {Sierra}}, \bibinfo {author}
  {\bibfnamefont {D.}~\bibnamefont {Nordlund}}, \bibinfo {author}
  {\bibfnamefont {C.~Y.}\ \bibnamefont {Hampton}}, \bibinfo {author}
  {\bibfnamefont {D.}~\bibnamefont {Starodub}}, \bibinfo {author}
  {\bibfnamefont {D.~P.}\ \bibnamefont {DePonte}}, \bibinfo {author}
  {\bibfnamefont {M.}~\bibnamefont {Beye}}, \bibinfo {author} {\bibfnamefont
  {C.}~\bibnamefont {Chen}}, \bibinfo {author} {\bibfnamefont {A.~V.}\
  \bibnamefont {Martin}}, \bibinfo {author} {\bibfnamefont {A.}~\bibnamefont
  {Barty}}, \bibinfo {author} {\bibfnamefont {K.~T.}\ \bibnamefont {Wikfeldt}},
  \bibinfo {author} {\bibfnamefont {T.~M.}\ \bibnamefont {Weiss}}, \bibinfo
  {author} {\bibfnamefont {C.}~\bibnamefont {Caronna}}, \bibinfo {author}
  {\bibfnamefont {J.}~\bibnamefont {Feldkamp}}, \bibinfo {author}
  {\bibfnamefont {L.~B.}\ \bibnamefont {Skinner}}, \bibinfo {author}
  {\bibfnamefont {M.~M.}\ \bibnamefont {Seibert}}, \bibinfo {author}
  {\bibfnamefont {M.}~\bibnamefont {Messerschmidt}}, \bibinfo {author}
  {\bibfnamefont {G.~J.}\ \bibnamefont {Williams}}, \bibinfo {author}
  {\bibfnamefont {S.}~\bibnamefont {Boutet}}, \bibinfo {author} {\bibfnamefont
  {L.~G.~M.}\ \bibnamefont {Pettersson}}, \bibinfo {author} {\bibfnamefont
  {M.~J.}\ \bibnamefont {Bogan}}, \ and\ \bibinfo {author} {\bibfnamefont
  {A.}~\bibnamefont {Nilsson}},\ }\href@noop {} {\bibfield  {journal} {\bibinfo
   {journal} {Nature}\ }\textbf {\bibinfo {volume} {510}},\ \bibinfo {pages}
  {381} (\bibinfo {year} {2014})}\BibitemShut {NoStop}%
\bibitem [{\citenamefont {Debenedetti}, \citenamefont {Sciortino},\ and\
  \citenamefont {Zerze}(2020)}]{debenedetti2020}%
  \BibitemOpen
  \bibfield  {author} {\bibinfo {author} {\bibfnamefont {P.~G.}\ \bibnamefont
  {Debenedetti}}, \bibinfo {author} {\bibfnamefont {F.}~\bibnamefont
  {Sciortino}}, \ and\ \bibinfo {author} {\bibfnamefont {G.~H.}\ \bibnamefont
  {Zerze}},\ }\href@noop {} {\bibfield  {journal} {\bibinfo  {journal}
  {Science}\ }\textbf {\bibinfo {volume} {369}},\ \bibinfo {pages} {289}
  (\bibinfo {year} {2020})}\BibitemShut {NoStop}%
\bibitem [{\citenamefont {Kringle}\ \emph {et~al.}(2020)\citenamefont
  {Kringle}, \citenamefont {Thornley}, \citenamefont {Kay},\ and\ \citenamefont
  {Kimmel}}]{KimmelReversible2020}%
  \BibitemOpen
  \bibfield  {author} {\bibinfo {author} {\bibfnamefont {L.}~\bibnamefont
  {Kringle}}, \bibinfo {author} {\bibfnamefont {W.~A.}\ \bibnamefont
  {Thornley}}, \bibinfo {author} {\bibfnamefont {B.~D.}\ \bibnamefont {Kay}}, \
  and\ \bibinfo {author} {\bibfnamefont {G.~A.}\ \bibnamefont {Kimmel}},\
  }\href@noop {} {\bibfield  {journal} {\bibinfo  {journal} {Science}\ }\textbf
  {\bibinfo {volume} {369}},\ \bibinfo {pages} {1490} (\bibinfo {year}
  {2020})}\BibitemShut {NoStop}%
\bibitem [{\citenamefont {Kim}\ \emph {et~al.}(2020)\citenamefont {Kim},
  \citenamefont {Amann-Winkel}, \citenamefont {Giovambattista}, \citenamefont
  {Spah}, \citenamefont {Perakis}, \citenamefont {Pathak}, \citenamefont
  {Parada}, \citenamefont {Yang}, \citenamefont {Mariedahl}, \citenamefont
  {Eklund}, \citenamefont {Lane}, \citenamefont {You}, \citenamefont {Jeong},
  \citenamefont {Weston}, \citenamefont {Lee}, \citenamefont {Eom},
  \citenamefont {Kim}, \citenamefont {Park}, \citenamefont {S.H.~Chun},\ and\
  \citenamefont {Nilsson}}]{experimental2020}%
  \BibitemOpen
  \bibfield  {author} {\bibinfo {author} {\bibfnamefont {K.~H.}\ \bibnamefont
  {Kim}}, \bibinfo {author} {\bibfnamefont {K.}~\bibnamefont {Amann-Winkel}},
  \bibinfo {author} {\bibfnamefont {N.}~\bibnamefont {Giovambattista}},
  \bibinfo {author} {\bibfnamefont {A.}~\bibnamefont {Spah}}, \bibinfo {author}
  {\bibfnamefont {F.}~\bibnamefont {Perakis}}, \bibinfo {author} {\bibfnamefont
  {H.}~\bibnamefont {Pathak}}, \bibinfo {author} {\bibfnamefont {M.~L.}\
  \bibnamefont {Parada}}, \bibinfo {author} {\bibfnamefont {C.}~\bibnamefont
  {Yang}}, \bibinfo {author} {\bibfnamefont {D.}~\bibnamefont {Mariedahl}},
  \bibinfo {author} {\bibfnamefont {T.}~\bibnamefont {Eklund}}, \bibinfo
  {author} {\bibfnamefont {T.~J.}\ \bibnamefont {Lane}}, \bibinfo {author}
  {\bibfnamefont {S.}~\bibnamefont {You}}, \bibinfo {author} {\bibfnamefont
  {S.}~\bibnamefont {Jeong}}, \bibinfo {author} {\bibfnamefont
  {M.}~\bibnamefont {Weston}}, \bibinfo {author} {\bibfnamefont {J.~H.}\
  \bibnamefont {Lee}}, \bibinfo {author} {\bibfnamefont {I.}~\bibnamefont
  {Eom}}, \bibinfo {author} {\bibfnamefont {M.}~\bibnamefont {Kim}}, \bibinfo
  {author} {\bibfnamefont {J.}~\bibnamefont {Park}}, \bibinfo {author}
  {\bibfnamefont {P.~P.}\ \bibnamefont {S.H.~Chun}}, \ and\ \bibinfo {author}
  {\bibfnamefont {A.}~\bibnamefont {Nilsson}},\ }\href@noop {} {\bibfield
  {journal} {\bibinfo  {journal} {Science}\ }\textbf {\bibinfo {volume}
  {370}},\ \bibinfo {pages} {978} (\bibinfo {year} {2020})}\BibitemShut
  {NoStop}%
\bibitem [{\citenamefont {Poole}\ \emph {et~al.}(1992)\citenamefont {Poole},
  \citenamefont {Sciortino}, \citenamefont {Essmann},\ and\ \citenamefont
  {Stanley}}]{poole_nature}%
  \BibitemOpen
  \bibfield  {author} {\bibinfo {author} {\bibfnamefont {P.~H.}\ \bibnamefont
  {Poole}}, \bibinfo {author} {\bibfnamefont {F.}~\bibnamefont {Sciortino}},
  \bibinfo {author} {\bibfnamefont {U.}~\bibnamefont {Essmann}}, \ and\
  \bibinfo {author} {\bibfnamefont {H.~E.}\ \bibnamefont {Stanley}},\
  }\href@noop {} {\bibfield  {journal} {\bibinfo  {journal} {Nature}\ }\textbf
  {\bibinfo {volume} {360}},\ \bibinfo {pages} {324} (\bibinfo {year}
  {1992})}\BibitemShut {NoStop}%
\bibitem [{\citenamefont {Liu}\ \emph {et~al.}(2010)\citenamefont {Liu},
  \citenamefont {Palmer}, \citenamefont {Panagiotopoulos},\ and\ \citenamefont
  {Debenedetti}}]{liu_2010}%
  \BibitemOpen
  \bibfield  {author} {\bibinfo {author} {\bibfnamefont {Y.}~\bibnamefont
  {Liu}}, \bibinfo {author} {\bibfnamefont {J.~C.}\ \bibnamefont {Palmer}},
  \bibinfo {author} {\bibfnamefont {A.~Z.}\ \bibnamefont {Panagiotopoulos}}, \
  and\ \bibinfo {author} {\bibfnamefont {P.~G.}\ \bibnamefont {Debenedetti}},\
  }\href@noop {} {\bibfield  {journal} {\bibinfo  {journal} {J. Chem. Phys.}\
  }\textbf {\bibinfo {volume} {137}},\ \bibinfo {pages} {214505} (\bibinfo
  {year} {2010})}\BibitemShut {NoStop}%
\bibitem [{\citenamefont {Limmer}\ and\ \citenamefont
  {Chandler}(2011)}]{limmer1}%
  \BibitemOpen
  \bibfield  {author} {\bibinfo {author} {\bibfnamefont {D.~T.}\ \bibnamefont
  {Limmer}}\ and\ \bibinfo {author} {\bibfnamefont {D.}~\bibnamefont
  {Chandler}},\ }\href@noop {} {\bibfield  {journal} {\bibinfo  {journal} {J.
  Chem. Phys.}\ }\textbf {\bibinfo {volume} {135}},\ \bibinfo {pages} {134503}
  (\bibinfo {year} {2011})}\BibitemShut {NoStop}%
\bibitem [{\citenamefont {Wikfeldt}, \citenamefont {Nilsson},\ and\
  \citenamefont {Pettersson}(2011)}]{wikfeldt_2011}%
  \BibitemOpen
  \bibfield  {author} {\bibinfo {author} {\bibfnamefont {K.~T.}\ \bibnamefont
  {Wikfeldt}}, \bibinfo {author} {\bibfnamefont {A.}~\bibnamefont {Nilsson}}, \
  and\ \bibinfo {author} {\bibfnamefont {L.~G.~M.}\ \bibnamefont
  {Pettersson}},\ }\href@noop {} {\bibfield  {journal} {\bibinfo  {journal}
  {Phys. Chem. Chem. Phys.}\ }\textbf {\bibinfo {volume} {13}},\ \bibinfo
  {pages} {19918} (\bibinfo {year} {2011})}\BibitemShut {NoStop}%
\bibitem [{\citenamefont {Palmer}, \citenamefont {Car},\ and\ \citenamefont
  {Debenedetti}(2013)}]{palmer_2013}%
  \BibitemOpen
  \bibfield  {author} {\bibinfo {author} {\bibfnamefont {J.~C.}\ \bibnamefont
  {Palmer}}, \bibinfo {author} {\bibfnamefont {R.}~\bibnamefont {Car}}, \ and\
  \bibinfo {author} {\bibfnamefont {P.~G.}\ \bibnamefont {Debenedetti}},\
  }\href@noop {} {\bibfield  {journal} {\bibinfo  {journal} {Faraday Discuss.}\
  }\textbf {\bibinfo {volume} {167}},\ \bibinfo {pages} {77} (\bibinfo {year}
  {2013})}\BibitemShut {NoStop}%
\bibitem [{\citenamefont {Limmer}\ and\ \citenamefont
  {Chandler}(2013)}]{limmer2}%
  \BibitemOpen
  \bibfield  {author} {\bibinfo {author} {\bibfnamefont {D.~T.}\ \bibnamefont
  {Limmer}}\ and\ \bibinfo {author} {\bibfnamefont {D.}~\bibnamefont
  {Chandler}},\ }\href@noop {} {\bibfield  {journal} {\bibinfo  {journal} {J.
  Chem. Phys.}\ }\textbf {\bibinfo {volume} {138}},\ \bibinfo {pages} {214504}
  (\bibinfo {year} {2013})}\BibitemShut {NoStop}%
\bibitem [{\citenamefont {Limmer}\ and\ \citenamefont
  {Chandler}(2014)}]{limmer3}%
  \BibitemOpen
  \bibfield  {author} {\bibinfo {author} {\bibfnamefont {D.~T.}\ \bibnamefont
  {Limmer}}\ and\ \bibinfo {author} {\bibfnamefont {D.}~\bibnamefont
  {Chandler}},\ }\href@noop {} {\bibfield  {journal} {\bibinfo  {journal}
  {Proc. Natl. Acad. Sci. USA}\ }\textbf {\bibinfo {volume} {111}},\ \bibinfo
  {pages} {9413} (\bibinfo {year} {2014})}\BibitemShut {NoStop}%
\bibitem [{\citenamefont {Chandler}(2016)}]{chandler_rosica}%
  \BibitemOpen
  \bibfield  {author} {\bibinfo {author} {\bibfnamefont {D.}~\bibnamefont
  {Chandler}},\ }\href@noop {} {\bibfield  {journal} {\bibinfo  {journal}
  {Nature}\ }\textbf {\bibinfo {volume} {531}},\ \bibinfo {pages} {E1}
  (\bibinfo {year} {2016})}\BibitemShut {NoStop}%
\bibitem [{\citenamefont {Palmer}\ \emph
  {et~al.}(2016{\natexlab{a}})\citenamefont {Palmer}, \citenamefont {Martelli},
  \citenamefont {Liu}, \citenamefont {Car}, \citenamefont {Panagiotopoulos},\
  and\ \citenamefont {P}}]{martelli_comment2}%
  \BibitemOpen
  \bibfield  {author} {\bibinfo {author} {\bibfnamefont {J.~C.}\ \bibnamefont
  {Palmer}}, \bibinfo {author} {\bibfnamefont {F.}~\bibnamefont {Martelli}},
  \bibinfo {author} {\bibfnamefont {Y.}~\bibnamefont {Liu}}, \bibinfo {author}
  {\bibfnamefont {R.}~\bibnamefont {Car}}, \bibinfo {author} {\bibfnamefont
  {A.~Z.}\ \bibnamefont {Panagiotopoulos}}, \ and\ \bibinfo {author}
  {\bibfnamefont {G.~D.}\ \bibnamefont {P}},\ }\href@noop {} {\bibfield
  {journal} {\bibinfo  {journal} {Nature}\ }\textbf {\bibinfo {volume} {531}},\
  \bibinfo {pages} {E2} (\bibinfo {year} {2016}{\natexlab{a}})}\BibitemShut
  {NoStop}%
\bibitem [{\citenamefont {Palmer}\ \emph {et~al.}(2018)\citenamefont {Palmer},
  \citenamefont {Haji-Akbari}, \citenamefont {Singh}, \citenamefont {Martelli},
  \citenamefont {Car}, \citenamefont {Panagiotopoulos},\ and\ \citenamefont
  {Debenedetti}}]{martelli_comment}%
  \BibitemOpen
  \bibfield  {author} {\bibinfo {author} {\bibfnamefont {J.~C.}\ \bibnamefont
  {Palmer}}, \bibinfo {author} {\bibfnamefont {A.}~\bibnamefont {Haji-Akbari}},
  \bibinfo {author} {\bibfnamefont {R.~S.}\ \bibnamefont {Singh}}, \bibinfo
  {author} {\bibfnamefont {F.}~\bibnamefont {Martelli}}, \bibinfo {author}
  {\bibfnamefont {R.}~\bibnamefont {Car}}, \bibinfo {author} {\bibfnamefont
  {A.~Z.}\ \bibnamefont {Panagiotopoulos}}, \ and\ \bibinfo {author}
  {\bibfnamefont {P.~G.}\ \bibnamefont {Debenedetti}},\ }\href@noop {}
  {\bibfield  {journal} {\bibinfo  {journal} {J. Chem. Phys.}\ }\textbf
  {\bibinfo {volume} {148}},\ \bibinfo {pages} {137101} (\bibinfo {year}
  {2018})}\BibitemShut {NoStop}%
\bibitem [{\citenamefont {Palmer}\ \emph
  {et~al.}(2016{\natexlab{b}})\citenamefont {Palmer}, \citenamefont {Singh},
  \citenamefont {Chen}, \citenamefont {Martelli},\ and\ \citenamefont
  {Debenedetti}}]{martelli_density}%
  \BibitemOpen
  \bibfield  {author} {\bibinfo {author} {\bibfnamefont {J.~C.}\ \bibnamefont
  {Palmer}}, \bibinfo {author} {\bibfnamefont {R.~S.}\ \bibnamefont {Singh}},
  \bibinfo {author} {\bibfnamefont {R.}~\bibnamefont {Chen}}, \bibinfo {author}
  {\bibfnamefont {F.}~\bibnamefont {Martelli}}, \ and\ \bibinfo {author}
  {\bibfnamefont {P.~G.}\ \bibnamefont {Debenedetti}},\ }\href@noop {}
  {\bibfield  {journal} {\bibinfo  {journal} {Mol. Phys.}\ }\textbf {\bibinfo
  {volume} {114}},\ \bibinfo {pages} {2580} (\bibinfo {year}
  {2016}{\natexlab{b}})}\BibitemShut {NoStop}%
\bibitem [{\citenamefont {Martelli}\ \emph {et~al.}(2020)\citenamefont
  {Martelli}, \citenamefont {Leoni}, \citenamefont {Sciortino},\ and\
  \citenamefont {Russo}}]{martelli_connection}%
  \BibitemOpen
  \bibfield  {author} {\bibinfo {author} {\bibfnamefont {F.}~\bibnamefont
  {Martelli}}, \bibinfo {author} {\bibfnamefont {F.}~\bibnamefont {Leoni}},
  \bibinfo {author} {\bibfnamefont {F.}~\bibnamefont {Sciortino}}, \ and\
  \bibinfo {author} {\bibfnamefont {J.}~\bibnamefont {Russo}},\ }\href@noop {}
  {\bibfield  {journal} {\bibinfo  {journal} {J. Chem. Phys.}\ }\textbf
  {\bibinfo {volume} {153}},\ \bibinfo {pages} {104503} (\bibinfo {year}
  {2020})}\BibitemShut {NoStop}%
\bibitem [{\citenamefont {Shi}\ and\ \citenamefont
  {Tanaka}(2020{\natexlab{a}})}]{ShiAnomalies2020}%
  \BibitemOpen
  \bibfield  {author} {\bibinfo {author} {\bibfnamefont {R.}~\bibnamefont
  {Shi}}\ and\ \bibinfo {author} {\bibfnamefont {H.}~\bibnamefont {Tanaka}},\
  }\href@noop {} {\bibfield  {journal} {\bibinfo  {journal} {Proc. Natl. Ac.
  Sci.}\ }\textbf {\bibinfo {volume} {117}},\ \bibinfo {pages} {26591}
  (\bibinfo {year} {2020}{\natexlab{a}})}\BibitemShut {NoStop}%
\bibitem [{\citenamefont {Shi}\ and\ \citenamefont
  {Tanaka}(2020{\natexlab{b}})}]{ShiDirect2020}%
  \BibitemOpen
  \bibfield  {author} {\bibinfo {author} {\bibfnamefont {R.}~\bibnamefont
  {Shi}}\ and\ \bibinfo {author} {\bibfnamefont {H.}~\bibnamefont {Tanaka}},\
  }\href@noop {} {\bibfield  {journal} {\bibinfo  {journal} {J. Am. Chem.
  Soc.}\ }\textbf {\bibinfo {volume} {142}},\ \bibinfo {pages} {2868} (\bibinfo
  {year} {2020}{\natexlab{b}})}\BibitemShut {NoStop}%
\bibitem [{\citenamefont {Shi}, \citenamefont {Russo},\ and\ \citenamefont
  {Tanaka}(2018{\natexlab{a}})}]{ShiCommon2018}%
  \BibitemOpen
  \bibfield  {author} {\bibinfo {author} {\bibfnamefont {R.}~\bibnamefont
  {Shi}}, \bibinfo {author} {\bibfnamefont {J.}~\bibnamefont {Russo}}, \ and\
  \bibinfo {author} {\bibfnamefont {H.}~\bibnamefont {Tanaka}},\ }\href@noop {}
  {\bibfield  {journal} {\bibinfo  {journal} {J. Chem. Phys.}\ }\textbf
  {\bibinfo {volume} {149}},\ \bibinfo {pages} {224502} (\bibinfo {year}
  {2018}{\natexlab{a}})}\BibitemShut {NoStop}%
\bibitem [{\citenamefont {Russo}\ and\ \citenamefont
  {Tanaka}(2014{\natexlab{a}})}]{RussoUnderstanding2014}%
  \BibitemOpen
  \bibfield  {author} {\bibinfo {author} {\bibfnamefont {J.}~\bibnamefont
  {Russo}}\ and\ \bibinfo {author} {\bibfnamefont {H.}~\bibnamefont {Tanaka}},\
  }\href@noop {} {\bibfield  {journal} {\bibinfo  {journal} {Nat. Commun.}\
  }\textbf {\bibinfo {volume} {5}},\ \bibinfo {pages} {3556} (\bibinfo {year}
  {2014}{\natexlab{a}})}\BibitemShut {NoStop}%
\bibitem [{\citenamefont {Akahane}\ and\ \citenamefont
  {Tanaka}(2018)}]{RussoWater2018}%
  \BibitemOpen
  \bibfield  {author} {\bibinfo {author} {\bibfnamefont {J.~R.~K.}\
  \bibnamefont {Akahane}}\ and\ \bibinfo {author} {\bibfnamefont
  {H.}~\bibnamefont {Tanaka}},\ }\href@noop {} {\bibfield  {journal} {\bibinfo
  {journal} {Proc. Natl. Ac. Sci.}\ }\textbf {\bibinfo {volume} {115}},\
  \bibinfo {pages} {E3333} (\bibinfo {year} {2018})}\BibitemShut {NoStop}%
\bibitem [{\citenamefont {Santra}\ \emph {et~al.}(2015)\citenamefont {Santra},
  \citenamefont {{DiStasio Jr.}}, \citenamefont {Martelli},\ and\ \citenamefont
  {Car}}]{santra_2015}%
  \BibitemOpen
  \bibfield  {author} {\bibinfo {author} {\bibfnamefont {B.}~\bibnamefont
  {Santra}}, \bibinfo {author} {\bibfnamefont {R.~A.}\ \bibnamefont {{DiStasio
  Jr.}}}, \bibinfo {author} {\bibfnamefont {F.}~\bibnamefont {Martelli}}, \
  and\ \bibinfo {author} {\bibfnamefont {R.}~\bibnamefont {Car}},\ }\href@noop
  {} {\bibfield  {journal} {\bibinfo  {journal} {Mol. Phys.}\ }\textbf
  {\bibinfo {volume} {113}},\ \bibinfo {pages} {2829} (\bibinfo {year}
  {2015})}\BibitemShut {NoStop}%
\bibitem [{\citenamefont {Huang}\ \emph {et~al.}(2009)\citenamefont {Huang},
  \citenamefont {Wikfeldt}, \citenamefont {Tokushima}, \citenamefont
  {Nordlund}, \citenamefont {Harada}, \citenamefont {Bergmann}, \citenamefont
  {Niebuhr}, \citenamefont {Weiss}, \citenamefont {Horikawa}, \citenamefont
  {Leetmaa}, \citenamefont {Ljungberg}, \citenamefont {Takahashi},
  \citenamefont {Lenz}, \citenamefont {Ojam\"{a}e}, \citenamefont {Lyubartsev},
  \citenamefont {Shin}, \citenamefont {Pettersson},\ and\ \citenamefont
  {Nilsson}}]{huang_2009}%
  \BibitemOpen
  \bibfield  {author} {\bibinfo {author} {\bibfnamefont {C.}~\bibnamefont
  {Huang}}, \bibinfo {author} {\bibfnamefont {K.~T.}\ \bibnamefont {Wikfeldt}},
  \bibinfo {author} {\bibfnamefont {T.}~\bibnamefont {Tokushima}}, \bibinfo
  {author} {\bibfnamefont {D.}~\bibnamefont {Nordlund}}, \bibinfo {author}
  {\bibfnamefont {Y.}~\bibnamefont {Harada}}, \bibinfo {author} {\bibfnamefont
  {U.}~\bibnamefont {Bergmann}}, \bibinfo {author} {\bibfnamefont
  {M.}~\bibnamefont {Niebuhr}}, \bibinfo {author} {\bibfnamefont {T.~M.}\
  \bibnamefont {Weiss}}, \bibinfo {author} {\bibfnamefont {Y.}~\bibnamefont
  {Horikawa}}, \bibinfo {author} {\bibfnamefont {M.}~\bibnamefont {Leetmaa}},
  \bibinfo {author} {\bibfnamefont {M.~P.}\ \bibnamefont {Ljungberg}}, \bibinfo
  {author} {\bibfnamefont {O.}~\bibnamefont {Takahashi}}, \bibinfo {author}
  {\bibfnamefont {A.}~\bibnamefont {Lenz}}, \bibinfo {author} {\bibfnamefont
  {L.}~\bibnamefont {Ojam\"{a}e}}, \bibinfo {author} {\bibfnamefont {A.~P.}\
  \bibnamefont {Lyubartsev}}, \bibinfo {author} {\bibfnamefont
  {S.}~\bibnamefont {Shin}}, \bibinfo {author} {\bibfnamefont {L.~G.~M.}\
  \bibnamefont {Pettersson}}, \ and\ \bibinfo {author} {\bibfnamefont
  {A.}~\bibnamefont {Nilsson}},\ }\href@noop {} {\bibfield  {journal} {\bibinfo
   {journal} {Proc. Natl. Acad. Sci. USA}\ }\textbf {\bibinfo {volume} {106}},\
  \bibinfo {pages} {15214} (\bibinfo {year} {2009})}\BibitemShut {NoStop}%
\bibitem [{\citenamefont {Nilsson}\ and\ \citenamefont
  {Pettersson}(2015)}]{nilsson_2015}%
  \BibitemOpen
  \bibfield  {author} {\bibinfo {author} {\bibfnamefont {A.}~\bibnamefont
  {Nilsson}}\ and\ \bibinfo {author} {\bibfnamefont {L.~G.~M.}\ \bibnamefont
  {Pettersson}},\ }\href@noop {} {\bibfield  {journal} {\bibinfo  {journal}
  {Nat. Commun.}\ }\textbf {\bibinfo {volume} {6}},\ \bibinfo {pages} {8998}
  (\bibinfo {year} {2015})}\BibitemShut {NoStop}%
\bibitem [{\citenamefont {{De Marzio}}\ \emph {et~al.}(2017)\citenamefont {{De
  Marzio}}, \citenamefont {Camicasca}, \citenamefont {Rovere},\ and\
  \citenamefont {Gallo}}]{demarzio_2017}%
  \BibitemOpen
  \bibfield  {author} {\bibinfo {author} {\bibfnamefont {M.}~\bibnamefont {{De
  Marzio}}}, \bibinfo {author} {\bibfnamefont {G.}~\bibnamefont {Camicasca}},
  \bibinfo {author} {\bibfnamefont {M.}~\bibnamefont {Rovere}}, \ and\ \bibinfo
  {author} {\bibfnamefont {P.}~\bibnamefont {Gallo}},\ }\href@noop {}
  {\bibfield  {journal} {\bibinfo  {journal} {J. Chem. Phys.}\ }\textbf
  {\bibinfo {volume} {146}},\ \bibinfo {pages} {084502} (\bibinfo {year}
  {2017})}\BibitemShut {NoStop}%
\bibitem [{\citenamefont {Martelli}(2019)}]{martelli_unravelling}%
  \BibitemOpen
  \bibfield  {author} {\bibinfo {author} {\bibfnamefont {F.}~\bibnamefont
  {Martelli}},\ }\href@noop {} {\bibfield  {journal} {\bibinfo  {journal} {J.
  Chem. Phys.}\ }\textbf {\bibinfo {volume} {150}},\ \bibinfo {pages} {094506}
  (\bibinfo {year} {2019})}\BibitemShut {NoStop}%
\bibitem [{\citenamefont {Pekka}\ and\ \citenamefont
  {Lennart}(2001)}]{comparison_1}%
  \BibitemOpen
  \bibfield  {author} {\bibinfo {author} {\bibfnamefont {M.}~\bibnamefont
  {Pekka}}\ and\ \bibinfo {author} {\bibfnamefont {N.}~\bibnamefont
  {Lennart}},\ }\href@noop {} {\bibfield  {journal} {\bibinfo  {journal} {J.
  Phys. Chem. A}\ }\textbf {\bibinfo {volume} {105}},\ \bibinfo {pages} {9954}
  (\bibinfo {year} {2001})}\BibitemShut {NoStop}%
\bibitem [{\citenamefont {Mao}\ and\ \citenamefont
  {Zhang}(2012)}]{comparison_2}%
  \BibitemOpen
  \bibfield  {author} {\bibinfo {author} {\bibfnamefont {Y.}~\bibnamefont
  {Mao}}\ and\ \bibinfo {author} {\bibfnamefont {Y.}~\bibnamefont {Zhang}},\
  }\href@noop {} {\bibfield  {journal} {\bibinfo  {journal} {Chem. Phys.
  Lett.}\ }\textbf {\bibinfo {volume} {542}},\ \bibinfo {pages} {37} (\bibinfo
  {year} {2012})}\BibitemShut {NoStop}%
\bibitem [{\citenamefont {Lee}\ and\ \citenamefont {Kim}(2019)}]{comparison_3}%
  \BibitemOpen
  \bibfield  {author} {\bibinfo {author} {\bibfnamefont {S.~H.}\ \bibnamefont
  {Lee}}\ and\ \bibinfo {author} {\bibfnamefont {J.}~\bibnamefont {Kim}},\
  }\href@noop {} {\bibfield  {journal} {\bibinfo  {journal} {Mol. Phys.}\
  }\textbf {\bibinfo {volume} {117}},\ \bibinfo {pages} {1926} (\bibinfo {year}
  {2019})}\BibitemShut {NoStop}%
\bibitem [{\citenamefont {Jorgensen}\ \emph {et~al.}(1983)\citenamefont
  {Jorgensen}, \citenamefont {Chandrasekhar}, \citenamefont {Madura},
  \citenamefont {Impey},\ and\ \citenamefont
  {Klein}}]{comparison_4_tip3p_tip4p}%
  \BibitemOpen
  \bibfield  {author} {\bibinfo {author} {\bibfnamefont {W.~L.}\ \bibnamefont
  {Jorgensen}}, \bibinfo {author} {\bibfnamefont {J.}~\bibnamefont
  {Chandrasekhar}}, \bibinfo {author} {\bibfnamefont {J.~D.}\ \bibnamefont
  {Madura}}, \bibinfo {author} {\bibfnamefont {R.~W.}\ \bibnamefont {Impey}}, \
  and\ \bibinfo {author} {\bibfnamefont {M.~L.}\ \bibnamefont {Klein}},\
  }\href@noop {} {\bibfield  {journal} {\bibinfo  {journal} {J. Chem. Phys.}\
  }\textbf {\bibinfo {volume} {79}},\ \bibinfo {pages} {926} (\bibinfo {year}
  {1983})}\BibitemShut {NoStop}%
\bibitem [{\citenamefont {Harrach}\ and\ \citenamefont
  {Drossel}(2014)}]{comparison_5}%
  \BibitemOpen
  \bibfield  {author} {\bibinfo {author} {\bibfnamefont {M.~F.}\ \bibnamefont
  {Harrach}}\ and\ \bibinfo {author} {\bibfnamefont {B.}~\bibnamefont
  {Drossel}},\ }\href@noop {} {\bibfield  {journal} {\bibinfo  {journal} {J.
  Chem. Phys.}\ }\textbf {\bibinfo {volume} {140}},\ \bibinfo {pages} {174501}
  (\bibinfo {year} {2014})}\BibitemShut {NoStop}%
\bibitem [{\citenamefont {Gonz\'{a}lez}\ \emph {et~al.}(2010)\citenamefont
  {Gonz\'{a}lez}, \citenamefont {Noya}, \citenamefont {Vega},\ and\
  \citenamefont {Ses\'{e}}}]{comparison_6}%
  \BibitemOpen
  \bibfield  {author} {\bibinfo {author} {\bibfnamefont {B.~S.}\ \bibnamefont
  {Gonz\'{a}lez}}, \bibinfo {author} {\bibfnamefont {E.~G.}\ \bibnamefont
  {Noya}}, \bibinfo {author} {\bibfnamefont {C.}~\bibnamefont {Vega}}, \ and\
  \bibinfo {author} {\bibfnamefont {L.~M.}\ \bibnamefont {Ses\'{e}}},\
  }\href@noop {} {\bibfield  {journal} {\bibinfo  {journal} {J. Phys. Chem. B}\
  }\textbf {\bibinfo {volume} {114}},\ \bibinfo {pages} {2484} (\bibinfo {year}
  {2010})}\BibitemShut {NoStop}%
\bibitem [{\citenamefont {Zielkiewicz}(2005)}]{comparison_7}%
  \BibitemOpen
  \bibfield  {author} {\bibinfo {author} {\bibfnamefont {J.}~\bibnamefont
  {Zielkiewicz}},\ }\href@noop {} {\bibfield  {journal} {\bibinfo  {journal}
  {J. Chem. Phys.}\ }\textbf {\bibinfo {volume} {123}},\ \bibinfo {pages}
  {104501} (\bibinfo {year} {2005})}\BibitemShut {NoStop}%
\bibitem [{\citenamefont {Steinczinger}, \citenamefont {J\'{o}v\'{a}ri},\ and\
  \citenamefont {Pusztai}(2017)}]{comparison_8}%
  \BibitemOpen
  \bibfield  {author} {\bibinfo {author} {\bibfnamefont {Z.}~\bibnamefont
  {Steinczinger}}, \bibinfo {author} {\bibfnamefont {P.}~\bibnamefont
  {J\'{o}v\'{a}ri}}, \ and\ \bibinfo {author} {\bibfnamefont {L.}~\bibnamefont
  {Pusztai}},\ }\href@noop {} {\bibfield  {journal} {\bibinfo  {journal} {J.
  Mol. Phys.}\ }\textbf {\bibinfo {volume} {228}},\ \bibinfo {pages} {19}
  (\bibinfo {year} {2017})}\BibitemShut {NoStop}%
\bibitem [{\citenamefont {Dix}, \citenamefont {Lue},\ and\ \citenamefont
  {Carbone}(2018)}]{DixWhy2018}%
  \BibitemOpen
  \bibfield  {author} {\bibinfo {author} {\bibfnamefont {J.}~\bibnamefont
  {Dix}}, \bibinfo {author} {\bibfnamefont {L.}~\bibnamefont {Lue}}, \ and\
  \bibinfo {author} {\bibfnamefont {P.}~\bibnamefont {Carbone}},\ }\href@noop
  {} {\bibfield  {journal} {\bibinfo  {journal} {J. Comp. Chem.}\ }\textbf
  {\bibinfo {volume} {39}},\ \bibinfo {pages} {2051} (\bibinfo {year}
  {2018})}\BibitemShut {NoStop}%
\bibitem [{\citenamefont {Bernal}\ and\ \citenamefont
  {Fowler}(1933)}]{bernal_fowler}%
  \BibitemOpen
  \bibfield  {author} {\bibinfo {author} {\bibfnamefont {J.~D.}\ \bibnamefont
  {Bernal}}\ and\ \bibinfo {author} {\bibfnamefont {R.~H.}\ \bibnamefont
  {Fowler}},\ }\href@noop {} {\bibfield  {journal} {\bibinfo  {journal} {J.
  Chem. Phys.}\ }\textbf {\bibinfo {volume} {1}},\ \bibinfo {pages} {515}
  (\bibinfo {year} {1933})}\BibitemShut {NoStop}%
\bibitem [{\citenamefont {Tse}\ \emph {et~al.}(1999)\citenamefont {Tse},
  \citenamefont {Klug}, \citenamefont {Tulk}, \citenamefont {Swainson},
  \citenamefont {Svensson}, \citenamefont {Loong}, \citenamefont {Shpakov},
  \citenamefont {Belosludov}, \citenamefont {Belosludov},\ and\ \citenamefont
  {Kawazoe}}]{TseMechanism1999}%
  \BibitemOpen
  \bibfield  {author} {\bibinfo {author} {\bibfnamefont {J.~S.}\ \bibnamefont
  {Tse}}, \bibinfo {author} {\bibfnamefont {D.~D.}\ \bibnamefont {Klug}},
  \bibinfo {author} {\bibfnamefont {C.~A.}\ \bibnamefont {Tulk}}, \bibinfo
  {author} {\bibfnamefont {I.}~\bibnamefont {Swainson}}, \bibinfo {author}
  {\bibfnamefont {E.~C.}\ \bibnamefont {Svensson}}, \bibinfo {author}
  {\bibfnamefont {C.-K.}\ \bibnamefont {Loong}}, \bibinfo {author}
  {\bibfnamefont {V.}~\bibnamefont {Shpakov}}, \bibinfo {author} {\bibfnamefont
  {V.~R.}\ \bibnamefont {Belosludov}}, \bibinfo {author} {\bibfnamefont
  {R.~V.}\ \bibnamefont {Belosludov}}, \ and\ \bibinfo {author} {\bibfnamefont
  {Y.}~\bibnamefont {Kawazoe}},\ }\href@noop {} {\bibfield  {journal} {\bibinfo
   {journal} {Nature}\ }\textbf {\bibinfo {volume} {400}},\ \bibinfo {pages}
  {647} (\bibinfo {year} {1999})}\BibitemShut {NoStop}%
\bibitem [{\citenamefont {Marto\u{n}\'{a}k}, \citenamefont {Donadio},\ and\
  \citenamefont {Parrinello}(2004)}]{martonak_2004}%
  \BibitemOpen
  \bibfield  {author} {\bibinfo {author} {\bibfnamefont {R.}~\bibnamefont
  {Marto\u{n}\'{a}k}}, \bibinfo {author} {\bibfnamefont {D.}~\bibnamefont
  {Donadio}}, \ and\ \bibinfo {author} {\bibfnamefont {M.}~\bibnamefont
  {Parrinello}},\ }\href@noop {} {\bibfield  {journal} {\bibinfo  {journal}
  {Phys. Rev. Lett.}\ }\textbf {\bibinfo {volume} {92}},\ \bibinfo {pages}
  {225702} (\bibinfo {year} {2004})}\BibitemShut {NoStop}%
\bibitem [{\citenamefont {Marto\u{n}\'{a}k}, \citenamefont {Donadio},\ and\
  \citenamefont {Parrinello}(2005)}]{martonak_2005}%
  \BibitemOpen
  \bibfield  {author} {\bibinfo {author} {\bibfnamefont {R.}~\bibnamefont
  {Marto\u{n}\'{a}k}}, \bibinfo {author} {\bibfnamefont {D.}~\bibnamefont
  {Donadio}}, \ and\ \bibinfo {author} {\bibfnamefont {M.}~\bibnamefont
  {Parrinello}},\ }\href@noop {} {\bibfield  {journal} {\bibinfo  {journal} {J.
  Chem. Phys.}\ }\textbf {\bibinfo {volume} {122}},\ \bibinfo {pages} {134501}
  (\bibinfo {year} {2005})}\BibitemShut {NoStop}%
\bibitem [{\citenamefont {Shephard}\ \emph {et~al.}(2017)\citenamefont
  {Shephard}, \citenamefont {Ling}, \citenamefont {Sosso}, \citenamefont
  {Michaelides}, \citenamefont {Slater},\ and\ \citenamefont
  {Salzmann}}]{shephard2017high}%
  \BibitemOpen
  \bibfield  {author} {\bibinfo {author} {\bibfnamefont {J.~J.}\ \bibnamefont
  {Shephard}}, \bibinfo {author} {\bibfnamefont {S.}~\bibnamefont {Ling}},
  \bibinfo {author} {\bibfnamefont {G.}~\bibnamefont {Sosso}}, \bibinfo
  {author} {\bibfnamefont {A.}~\bibnamefont {Michaelides}}, \bibinfo {author}
  {\bibfnamefont {B.}~\bibnamefont {Slater}}, \ and\ \bibinfo {author}
  {\bibfnamefont {C.~G.}\ \bibnamefont {Salzmann}},\ }\href@noop {} {\bibfield
  {journal} {\bibinfo  {journal} {J. Phys. Chem. Lett.}\ }\textbf {\bibinfo
  {volume} {8}},\ \bibinfo {pages} {1645} (\bibinfo {year} {2017})}\BibitemShut
  {NoStop}%
\bibitem [{\citenamefont {Martelli}\ \emph {et~al.}(2018)\citenamefont
  {Martelli}, \citenamefont {Giovambattista}, \citenamefont {Torquato},\ and\
  \citenamefont {Car}}]{martelli_searching}%
  \BibitemOpen
  \bibfield  {author} {\bibinfo {author} {\bibfnamefont {F.}~\bibnamefont
  {Martelli}}, \bibinfo {author} {\bibfnamefont {N.}~\bibnamefont
  {Giovambattista}}, \bibinfo {author} {\bibfnamefont {S.}~\bibnamefont
  {Torquato}}, \ and\ \bibinfo {author} {\bibfnamefont {R.}~\bibnamefont
  {Car}},\ }\href@noop {} {\bibfield  {journal} {\bibinfo  {journal} {Phys.
  Rev. Materials}\ }\textbf {\bibinfo {volume} {2}},\ \bibinfo {pages} {075601}
  (\bibinfo {year} {2018})}\BibitemShut {NoStop}%
\bibitem [{\citenamefont {Martelli}, \citenamefont {Crain},\ and\ \citenamefont
  {Franzese}(2020)}]{martelli_acsnano}%
  \BibitemOpen
  \bibfield  {author} {\bibinfo {author} {\bibfnamefont {F.}~\bibnamefont
  {Martelli}}, \bibinfo {author} {\bibfnamefont {J.}~\bibnamefont {Crain}}, \
  and\ \bibinfo {author} {\bibfnamefont {G.}~\bibnamefont {Franzese}},\
  }\href@noop {} {\bibfield  {journal} {\bibinfo  {journal} {ACS Nano}\
  }\textbf {\bibinfo {volume} {14}},\ \bibinfo {pages} {8616} (\bibinfo {year}
  {2020})}\BibitemShut {NoStop}%
\bibitem [{\citenamefont {Chiricotto}\ \emph {et~al.}()\citenamefont
  {Chiricotto}, \citenamefont {Martelli}, \citenamefont {Giunta},\ and\
  \citenamefont {Carbone}}]{martelli_graphene}%
  \BibitemOpen
  \bibfield  {author} {\bibinfo {author} {\bibfnamefont {M.}~\bibnamefont
  {Chiricotto}}, \bibinfo {author} {\bibfnamefont {F.}~\bibnamefont
  {Martelli}}, \bibinfo {author} {\bibfnamefont {G.}~\bibnamefont {Giunta}}, \
  and\ \bibinfo {author} {\bibfnamefont {P.}~\bibnamefont {Carbone}},\
  }\href@noop {} {\bibfield  {journal} {\bibinfo  {journal} {J. Phys. Chem. C}\
  }}\bibinfo {note} {Under review}\BibitemShut {NoStop}%
\bibitem [{\citenamefont {Martelli}, \citenamefont {Calero},\ and\
  \citenamefont {Franzese}(2021)}]{martelli_redefining}%
  \BibitemOpen
  \bibfield  {author} {\bibinfo {author} {\bibfnamefont {F.}~\bibnamefont
  {Martelli}}, \bibinfo {author} {\bibfnamefont {C.}~\bibnamefont {Calero}}, \
  and\ \bibinfo {author} {\bibfnamefont {G.}~\bibnamefont {Franzese}},\
  }\href@noop {} {\enquote {\bibinfo {title} {Re-defining the concept of
  hydration water near soft interfaces},}\ } (\bibinfo {year} {2021}),\ \Eprint
  {http://arxiv.org/abs/2101.06136} {arXiv:2101.06136 [cond-mat.soft]}
  \BibitemShut {NoStop}%
\bibitem [{\citenamefont {Berendsen}\ \emph {et~al.}(1981)\citenamefont
  {Berendsen}, \citenamefont {Postma}, \citenamefont {van Gunsteren},\ and\
  \citenamefont {Hermans}}]{spc}%
  \BibitemOpen
  \bibfield  {author} {\bibinfo {author} {\bibfnamefont {H.~J.~C.}\
  \bibnamefont {Berendsen}}, \bibinfo {author} {\bibfnamefont {J.~P.~M.}\
  \bibnamefont {Postma}}, \bibinfo {author} {\bibfnamefont {W.~F.}\
  \bibnamefont {van Gunsteren}}, \ and\ \bibinfo {author} {\bibfnamefont
  {J.}~\bibnamefont {Hermans}},\ }in\ \href@noop {} {\emph {\bibinfo
  {booktitle} {Intermolecular Forces}}},\ \bibinfo {editor} {edited by\
  \bibinfo {editor} {\bibfnamefont {B.}~\bibnamefont {Pullman}}}\ (\bibinfo
  {publisher} {Springer},\ \bibinfo {address} {Dordrecht},\ \bibinfo {year}
  {1981})\ pp.\ \bibinfo {pages} {331--342}\BibitemShut {NoStop}%
\bibitem [{\citenamefont {Berendsen}, \citenamefont {Grigera},\ and\
  \citenamefont {Straatsma}(1987)}]{spce}%
  \BibitemOpen
  \bibfield  {author} {\bibinfo {author} {\bibfnamefont {H.~J.~C.}\
  \bibnamefont {Berendsen}}, \bibinfo {author} {\bibfnamefont {J.~R.}\
  \bibnamefont {Grigera}}, \ and\ \bibinfo {author} {\bibfnamefont {T.~P.}\
  \bibnamefont {Straatsma}},\ }\href@noop {} {\bibfield  {journal} {\bibinfo
  {journal} {J. Phys. Chem.}\ }\textbf {\bibinfo {volume} {91}},\ \bibinfo
  {pages} {6269} (\bibinfo {year} {1987})}\BibitemShut {NoStop}%
\bibitem [{\citenamefont {Toukan}\ and\ \citenamefont
  {Rahman}(1985)}]{spcflex}%
  \BibitemOpen
  \bibfield  {author} {\bibinfo {author} {\bibfnamefont {K.}~\bibnamefont
  {Toukan}}\ and\ \bibinfo {author} {\bibfnamefont {A.}~\bibnamefont
  {Rahman}},\ }\href@noop {} {\bibfield  {journal} {\bibinfo  {journal} {Phys.
  Rev. B}\ }\textbf {\bibinfo {volume} {31}},\ \bibinfo {pages} {2643}
  (\bibinfo {year} {1985})}\BibitemShut {NoStop}%
\bibitem [{\citenamefont {Amira}, \citenamefont {Sp\'{a}ngberg},\ and\
  \citenamefont {Hermansson}(2004)}]{spcflex2}%
  \BibitemOpen
  \bibfield  {author} {\bibinfo {author} {\bibfnamefont {S.}~\bibnamefont
  {Amira}}, \bibinfo {author} {\bibfnamefont {D.}~\bibnamefont
  {Sp\'{a}ngberg}}, \ and\ \bibinfo {author} {\bibfnamefont {K.}~\bibnamefont
  {Hermansson}},\ }\href@noop {} {\bibfield  {journal} {\bibinfo  {journal}
  {Chem. Phys.}\ }\textbf {\bibinfo {volume} {303}},\ \bibinfo {pages} {327}
  (\bibinfo {year} {2004})}\BibitemShut {NoStop}%
\bibitem [{\citenamefont {Abascal}\ \emph {et~al.}(2005)\citenamefont
  {Abascal}, \citenamefont {Sanz}, \citenamefont {Fernández},\ and\
  \citenamefont {Vega}}]{tip4pIce}%
  \BibitemOpen
  \bibfield  {author} {\bibinfo {author} {\bibfnamefont {J.~L.~F.}\
  \bibnamefont {Abascal}}, \bibinfo {author} {\bibfnamefont {E.}~\bibnamefont
  {Sanz}}, \bibinfo {author} {\bibfnamefont {R.~G.}\ \bibnamefont
  {Fernández}}, \ and\ \bibinfo {author} {\bibfnamefont {C.}~\bibnamefont
  {Vega}},\ }\href@noop {} {\bibfield  {journal} {\bibinfo  {journal} {J. Chem.
  Phys.}\ }\textbf {\bibinfo {volume} {122}},\ \bibinfo {pages} {234511}
  (\bibinfo {year} {2005})}\BibitemShut {NoStop}%
\bibitem [{\citenamefont {Abascal}\ and\ \citenamefont
  {Vega}(2005)}]{tip4p2005}%
  \BibitemOpen
  \bibfield  {author} {\bibinfo {author} {\bibfnamefont {J.~L.~F.}\
  \bibnamefont {Abascal}}\ and\ \bibinfo {author} {\bibfnamefont
  {C.}~\bibnamefont {Vega}},\ }\href@noop {} {\bibfield  {journal} {\bibinfo
  {journal} {J. Chem. Phys.}\ }\textbf {\bibinfo {volume} {123}},\ \bibinfo
  {pages} {234505} (\bibinfo {year} {2005})}\BibitemShut {NoStop}%
\bibitem [{\citenamefont {Gonz\'{a}lez}\ and\ \citenamefont
  {Abascala}(2011)}]{tip4p2005flex}%
  \BibitemOpen
  \bibfield  {author} {\bibinfo {author} {\bibfnamefont {M.~A.}\ \bibnamefont
  {Gonz\'{a}lez}}\ and\ \bibinfo {author} {\bibfnamefont {J.~L.~F.}\
  \bibnamefont {Abascala}},\ }\href@noop {} {\bibfield  {journal} {\bibinfo
  {journal} {J. Chem. Phys.}\ }\textbf {\bibinfo {volume} {135}},\ \bibinfo
  {pages} {224516} (\bibinfo {year} {2011})}\BibitemShut {NoStop}%
\bibitem [{\citenamefont {Horn}\ \emph {et~al.}(2004)\citenamefont {Horn},
  \citenamefont {Swope}, \citenamefont {Pitera}, \citenamefont {Madura},
  \citenamefont {Dick}, \citenamefont {Hura},\ and\ \citenamefont
  {Head-Gordon}}]{tip4pEw}%
  \BibitemOpen
  \bibfield  {author} {\bibinfo {author} {\bibfnamefont {H.~W.}\ \bibnamefont
  {Horn}}, \bibinfo {author} {\bibfnamefont {W.~C.}\ \bibnamefont {Swope}},
  \bibinfo {author} {\bibfnamefont {J.~W.}\ \bibnamefont {Pitera}}, \bibinfo
  {author} {\bibfnamefont {J.~D.}\ \bibnamefont {Madura}}, \bibinfo {author}
  {\bibfnamefont {T.~J.}\ \bibnamefont {Dick}}, \bibinfo {author}
  {\bibfnamefont {G.~L.}\ \bibnamefont {Hura}}, \ and\ \bibinfo {author}
  {\bibfnamefont {T.}~\bibnamefont {Head-Gordon}},\ }\href@noop {} {\bibfield
  {journal} {\bibinfo  {journal} {J. Chem. Phys.}\ }\textbf {\bibinfo {volume}
  {120}},\ \bibinfo {pages} {9665} (\bibinfo {year} {2004})}\BibitemShut
  {NoStop}%
\bibitem [{\citenamefont {Mahoney}\ and\ \citenamefont
  {Jorgensen}(2000)}]{tip5p}%
  \BibitemOpen
  \bibfield  {author} {\bibinfo {author} {\bibfnamefont {M.~W.}\ \bibnamefont
  {Mahoney}}\ and\ \bibinfo {author} {\bibfnamefont {W.~L.}\ \bibnamefont
  {Jorgensen}},\ }\href@noop {} {\bibfield  {journal} {\bibinfo  {journal} {J.
  Chem. Phys.}\ }\textbf {\bibinfo {volume} {122}},\ \bibinfo {pages} {8910}
  (\bibinfo {year} {2000})}\BibitemShut {NoStop}%
\bibitem [{\citenamefont {Rick}(2004)}]{tip5pEw}%
  \BibitemOpen
  \bibfield  {author} {\bibinfo {author} {\bibfnamefont {S.~W.}\ \bibnamefont
  {Rick}},\ }\href@noop {} {\bibfield  {journal} {\bibinfo  {journal} {J. Chem.
  Phys.}\ }\textbf {\bibinfo {volume} {120}},\ \bibinfo {pages} {6085}
  (\bibinfo {year} {2004})}\BibitemShut {NoStop}%
\bibitem [{\citenamefont {Martelli}\ \emph {et~al.}(2016)\citenamefont
  {Martelli}, \citenamefont {Ko}, \citenamefont {O\u{g}uz},\ and\ \citenamefont
  {Car}}]{martelli_LOM}%
  \BibitemOpen
  \bibfield  {author} {\bibinfo {author} {\bibfnamefont {F.}~\bibnamefont
  {Martelli}}, \bibinfo {author} {\bibfnamefont {H.-Y.}\ \bibnamefont {Ko}},
  \bibinfo {author} {\bibfnamefont {E.~C.}\ \bibnamefont {O\u{g}uz}}, \ and\
  \bibinfo {author} {\bibfnamefont {R.}~\bibnamefont {Car}},\ }\href@noop {}
  {\bibfield  {journal} {\bibinfo  {journal} {Phys. Rev. B}\ }\textbf {\bibinfo
  {volume} {97}},\ \bibinfo {pages} {064105} (\bibinfo {year}
  {2016})}\BibitemShut {NoStop}%
\bibitem [{\citenamefont {Formanek}\ and\ \citenamefont
  {Martelli}(2020)}]{martelli_rings}%
  \BibitemOpen
  \bibfield  {author} {\bibinfo {author} {\bibfnamefont {M.}~\bibnamefont
  {Formanek}}\ and\ \bibinfo {author} {\bibfnamefont {F.}~\bibnamefont
  {Martelli}},\ }\href@noop {} {\bibfield  {journal} {\bibinfo  {journal} {AIP
  Adv.}\ }\textbf {\bibinfo {volume} {10}},\ \bibinfo {pages} {055205}
  (\bibinfo {year} {2020})}\BibitemShut {NoStop}%
\bibitem [{\citenamefont {Leoni}\ \emph {et~al.}(2019)\citenamefont {Leoni},
  \citenamefont {Shi}, \citenamefont {Tanaka},\ and\ \citenamefont
  {Russo}}]{leoni_2019}%
  \BibitemOpen
  \bibfield  {author} {\bibinfo {author} {\bibfnamefont {F.}~\bibnamefont
  {Leoni}}, \bibinfo {author} {\bibfnamefont {R.}~\bibnamefont {Shi}}, \bibinfo
  {author} {\bibfnamefont {H.}~\bibnamefont {Tanaka}}, \ and\ \bibinfo {author}
  {\bibfnamefont {J.}~\bibnamefont {Russo}},\ }\href@noop {} {\bibfield
  {journal} {\bibinfo  {journal} {J. Chem. Phys.}\ }\textbf {\bibinfo {volume}
  {151}},\ \bibinfo {pages} {044505} (\bibinfo {year} {2019})}\BibitemShut
  {NoStop}%
\bibitem [{\citenamefont {Camisasca}\ \emph {et~al.}(2019)\citenamefont
  {Camisasca}, \citenamefont {Schlesinger}, \citenamefont {Zhovtobriukh},
  \citenamefont {Pitsevich},\ and\ \citenamefont
  {Pettersson}}]{camisasca_proposal}%
  \BibitemOpen
  \bibfield  {author} {\bibinfo {author} {\bibfnamefont {G.}~\bibnamefont
  {Camisasca}}, \bibinfo {author} {\bibfnamefont {D.}~\bibnamefont
  {Schlesinger}}, \bibinfo {author} {\bibfnamefont {I.}~\bibnamefont
  {Zhovtobriukh}}, \bibinfo {author} {\bibfnamefont {G.}~\bibnamefont
  {Pitsevich}}, \ and\ \bibinfo {author} {\bibfnamefont {L.~G.~M.}\
  \bibnamefont {Pettersson}},\ }\href@noop {} {\bibfield  {journal} {\bibinfo
  {journal} {J. Chem. Phys.}\ }\textbf {\bibinfo {volume} {151}},\ \bibinfo
  {pages} {034508} (\bibinfo {year} {2019})}\BibitemShut {NoStop}%
\bibitem [{\citenamefont {Russo}\ and\ \citenamefont
  {Tanaka}(2014{\natexlab{b}})}]{russo_2014}%
  \BibitemOpen
  \bibfield  {author} {\bibinfo {author} {\bibfnamefont {J.}~\bibnamefont
  {Russo}}\ and\ \bibinfo {author} {\bibfnamefont {H.}~\bibnamefont {Tanaka}},\
  }\href@noop {} {\bibfield  {journal} {\bibinfo  {journal} {Nat. Commun.}\
  }\textbf {\bibinfo {volume} {5}},\ \bibinfo {pages} {3556} (\bibinfo {year}
  {2014}{\natexlab{b}})}\BibitemShut {NoStop}%
\bibitem [{\citenamefont {Fitzner}\ \emph {et~al.}(2019)\citenamefont
  {Fitzner}, \citenamefont {Sosso}, \citenamefont {Cox},\ and\ \citenamefont
  {Michaelides}}]{fitzner_ice}%
  \BibitemOpen
  \bibfield  {author} {\bibinfo {author} {\bibfnamefont {M.}~\bibnamefont
  {Fitzner}}, \bibinfo {author} {\bibfnamefont {G.~C.}\ \bibnamefont {Sosso}},
  \bibinfo {author} {\bibfnamefont {S.~J.}\ \bibnamefont {Cox}}, \ and\
  \bibinfo {author} {\bibfnamefont {A.}~\bibnamefont {Michaelides}},\
  }\href@noop {} {\bibfield  {journal} {\bibinfo  {journal} {Proc. Natl. Acad.
  Sci. USA}\ }\textbf {\bibinfo {volume} {116}},\ \bibinfo {pages} {2009}
  (\bibinfo {year} {2019})}\BibitemShut {NoStop}%
\bibitem [{\citenamefont {Shi}\ and\ \citenamefont
  {Tanaka}(2018)}]{ShiImpact2018}%
  \BibitemOpen
  \bibfield  {author} {\bibinfo {author} {\bibfnamefont {R.}~\bibnamefont
  {Shi}}\ and\ \bibinfo {author} {\bibfnamefont {H.}~\bibnamefont {Tanaka}},\
  }\href@noop {} {\bibfield  {journal} {\bibinfo  {journal} {Proc. Natl. Acad.
  Sci. USA}\ }\textbf {\bibinfo {volume} {115}},\ \bibinfo {pages} {1980}
  (\bibinfo {year} {2018})}\BibitemShut {NoStop}%
\bibitem [{\citenamefont {Bak\'{o}}\ \emph {et~al.}(2017)\citenamefont
  {Bak\'{o}}, \citenamefont {Ol\'{a}h}, \citenamefont {L\'{a}bas},
  \citenamefont {B\'{a}lint}, \citenamefont {Pusztai},\ and\ \citenamefont
  {Funel}}]{rings_solutions1}%
  \BibitemOpen
  \bibfield  {author} {\bibinfo {author} {\bibfnamefont {I.}~\bibnamefont
  {Bak\'{o}}}, \bibinfo {author} {\bibfnamefont {J.}~\bibnamefont {Ol\'{a}h}},
  \bibinfo {author} {\bibfnamefont {A.}~\bibnamefont {L\'{a}bas}}, \bibinfo
  {author} {\bibfnamefont {S.}~\bibnamefont {B\'{a}lint}}, \bibinfo {author}
  {\bibfnamefont {L.}~\bibnamefont {Pusztai}}, \ and\ \bibinfo {author}
  {\bibfnamefont {M.~C.~B.}\ \bibnamefont {Funel}},\ }\href@noop {} {\bibfield
  {journal} {\bibinfo  {journal} {J. Mol. Liq.}\ }\textbf {\bibinfo {volume}
  {228}},\ \bibinfo {pages} {25} (\bibinfo {year} {2017})}\BibitemShut
  {NoStop}%
\bibitem [{\citenamefont {Pothoczki}, \citenamefont {Pusztai},\ and\
  \citenamefont {Bak\'{o}}(2018)}]{rings_solutions2}%
  \BibitemOpen
  \bibfield  {author} {\bibinfo {author} {\bibfnamefont {S.}~\bibnamefont
  {Pothoczki}}, \bibinfo {author} {\bibfnamefont {L.}~\bibnamefont {Pusztai}},
  \ and\ \bibinfo {author} {\bibfnamefont {I.}~\bibnamefont {Bak\'{o}}},\
  }\href@noop {} {\bibfield  {journal} {\bibinfo  {journal} {J. Phys. Chem. B}\
  }\textbf {\bibinfo {volume} {122}},\ \bibinfo {pages} {6790} (\bibinfo {year}
  {2018})}\BibitemShut {NoStop}%
\bibitem [{\citenamefont {Pothoczki}, \citenamefont {Pusztai},\ and\
  \citenamefont {Bak\'{o}}(2019)}]{rings_solutions3}%
  \BibitemOpen
  \bibfield  {author} {\bibinfo {author} {\bibfnamefont {S.}~\bibnamefont
  {Pothoczki}}, \bibinfo {author} {\bibfnamefont {L.}~\bibnamefont {Pusztai}},
  \ and\ \bibinfo {author} {\bibfnamefont {I.}~\bibnamefont {Bak\'{o}}},\
  }\href@noop {} {\bibfield  {journal} {\bibinfo  {journal} {J. Phys. Chem. B}\
  }\textbf {\bibinfo {volume} {123}},\ \bibinfo {pages} {7599} (\bibinfo {year}
  {2019})}\BibitemShut {NoStop}%
\bibitem [{\citenamefont {Li}\ \emph {et~al.}(2020{\natexlab{a}})\citenamefont
  {Li}, \citenamefont {Zhong}, \citenamefont {Yan}, \citenamefont {Zhang},
  \citenamefont {Xu}, \citenamefont {Francisco},\ and\ \citenamefont
  {Zeng}}]{LiUnraveling2020}%
  \BibitemOpen
  \bibfield  {author} {\bibinfo {author} {\bibfnamefont {L.}~\bibnamefont
  {Li}}, \bibinfo {author} {\bibfnamefont {J.}~\bibnamefont {Zhong}}, \bibinfo
  {author} {\bibfnamefont {Y.}~\bibnamefont {Yan}}, \bibinfo {author}
  {\bibfnamefont {J.}~\bibnamefont {Zhang}}, \bibinfo {author} {\bibfnamefont
  {J.}~\bibnamefont {Xu}}, \bibinfo {author} {\bibfnamefont {J.~S.}\
  \bibnamefont {Francisco}}, \ and\ \bibinfo {author} {\bibfnamefont {X.~C.}\
  \bibnamefont {Zeng}},\ }\href@noop {} {\bibfield  {journal} {\bibinfo
  {journal} {Proc. Natl. Acad. Sci. USA}\ }\textbf {\bibinfo {volume} {117}},\
  \bibinfo {pages} {24701} (\bibinfo {year} {2020}{\natexlab{a}})}\BibitemShut
  {NoStop}%
\bibitem [{\citenamefont {{DiStasio Jr.}}\ \emph {et~al.}(2014)\citenamefont
  {{DiStasio Jr.}}, \citenamefont {Santra}, \citenamefont {Li}, \citenamefont
  {Wu},\ and\ \citenamefont {Car}}]{distasio_2014}%
  \BibitemOpen
  \bibfield  {author} {\bibinfo {author} {\bibfnamefont {R.~A.}\ \bibnamefont
  {{DiStasio Jr.}}}, \bibinfo {author} {\bibfnamefont {B.}~\bibnamefont
  {Santra}}, \bibinfo {author} {\bibfnamefont {Z.}~\bibnamefont {Li}}, \bibinfo
  {author} {\bibfnamefont {X.}~\bibnamefont {Wu}}, \ and\ \bibinfo {author}
  {\bibfnamefont {R.}~\bibnamefont {Car}},\ }\href@noop {} {\bibfield
  {journal} {\bibinfo  {journal} {J. Chem. Phys.}\ }\textbf {\bibinfo {volume}
  {141}},\ \bibinfo {pages} {084502} (\bibinfo {year} {2014})}\BibitemShut
  {NoStop}%
\bibitem [{\citenamefont {Nos\'e}(1984)}]{nose}%
  \BibitemOpen
  \bibfield  {author} {\bibinfo {author} {\bibfnamefont {S.}~\bibnamefont
  {Nos\'e}},\ }\href@noop {} {\bibfield  {journal} {\bibinfo  {journal} {Mol.
  Phys.}\ }\textbf {\bibinfo {volume} {52}},\ \bibinfo {pages} {255} (\bibinfo
  {year} {1984})}\BibitemShut {NoStop}%
\bibitem [{\citenamefont {Hoover}(1985)}]{hoover}%
  \BibitemOpen
  \bibfield  {author} {\bibinfo {author} {\bibfnamefont {W.~G.}\ \bibnamefont
  {Hoover}},\ }\href@noop {} {\bibfield  {journal} {\bibinfo  {journal} {Phys.
  Rev. A}\ }\textbf {\bibinfo {volume} {31}},\ \bibinfo {pages} {1695}
  (\bibinfo {year} {1985})}\BibitemShut {NoStop}%
\bibitem [{\citenamefont {Parrinello}\ and\ \citenamefont
  {Rahman}(1981)}]{parrinello_rahman}%
  \BibitemOpen
  \bibfield  {author} {\bibinfo {author} {\bibfnamefont {M.}~\bibnamefont
  {Parrinello}}\ and\ \bibinfo {author} {\bibfnamefont {A.}~\bibnamefont
  {Rahman}},\ }\href@noop {} {\bibfield  {journal} {\bibinfo  {journal} {J.
  Appl. Phys.}\ }\textbf {\bibinfo {volume} {52}},\ \bibinfo {pages} {7182}
  (\bibinfo {year} {1981})}\BibitemShut {NoStop}%
\bibitem [{\citenamefont {Abraham}\ \emph {et~al.}(2015)\citenamefont
  {Abraham}, \citenamefont {Murtola}, \citenamefont {Schulz}, \citenamefont
  {P\'all}, \citenamefont {Smith}, \citenamefont {Hess},\ and\ \citenamefont
  {Lindahl}}]{gromacs}%
  \BibitemOpen
  \bibfield  {author} {\bibinfo {author} {\bibfnamefont {M.~J.}\ \bibnamefont
  {Abraham}}, \bibinfo {author} {\bibfnamefont {T.}~\bibnamefont {Murtola}},
  \bibinfo {author} {\bibfnamefont {R.}~\bibnamefont {Schulz}}, \bibinfo
  {author} {\bibfnamefont {S.}~\bibnamefont {P\'all}}, \bibinfo {author}
  {\bibfnamefont {J.~C.}\ \bibnamefont {Smith}}, \bibinfo {author}
  {\bibfnamefont {B.}~\bibnamefont {Hess}}, \ and\ \bibinfo {author}
  {\bibfnamefont {E.}~\bibnamefont {Lindahl}},\ }\href@noop {} {\bibfield
  {journal} {\bibinfo  {journal} {SoftwareX}\ }\textbf {\bibinfo {volume}
  {1}},\ \bibinfo {pages} {19} (\bibinfo {year} {2015})}\BibitemShut {NoStop}%
\bibitem [{\citenamefont {King}(1967)}]{king}%
  \BibitemOpen
  \bibfield  {author} {\bibinfo {author} {\bibfnamefont {S.~V.}\ \bibnamefont
  {King}},\ }\href@noop {} {\bibfield  {journal} {\bibinfo  {journal} {Nature}\
  }\textbf {\bibinfo {volume} {213}},\ \bibinfo {pages} {1112} (\bibinfo {year}
  {1967})}\BibitemShut {NoStop}%
\bibitem [{\citenamefont {Rahman}\ and\ \citenamefont
  {Stillinger}(1973)}]{rahman_hydrogen}%
  \BibitemOpen
  \bibfield  {author} {\bibinfo {author} {\bibfnamefont {A.}~\bibnamefont
  {Rahman}}\ and\ \bibinfo {author} {\bibfnamefont {F.~H.}\ \bibnamefont
  {Stillinger}},\ }\href@noop {} {\bibfield  {journal} {\bibinfo  {journal} {J.
  Am. Chem. Soc.}\ }\textbf {\bibinfo {volume} {95}},\ \bibinfo {pages} {7943}
  (\bibinfo {year} {1973})}\BibitemShut {NoStop}%
\bibitem [{\citenamefont {Guttman}(1990)}]{guttman_ring}%
  \BibitemOpen
  \bibfield  {author} {\bibinfo {author} {\bibfnamefont {L.}~\bibnamefont
  {Guttman}},\ }\href@noop {} {\bibfield  {journal} {\bibinfo  {journal} {J.
  Non-Cryst. Solids}\ }\textbf {\bibinfo {volume} {116}},\ \bibinfo {pages}
  {145} (\bibinfo {year} {1990})}\BibitemShut {NoStop}%
\bibitem [{\citenamefont {Franzblau}(1991)}]{franzblau_computation}%
  \BibitemOpen
  \bibfield  {author} {\bibinfo {author} {\bibfnamefont {D.~S.}\ \bibnamefont
  {Franzblau}},\ }\href@noop {} {\bibfield  {journal} {\bibinfo  {journal}
  {Phys. Rev. B}\ }\textbf {\bibinfo {volume} {44}},\ \bibinfo {pages} {4925}
  (\bibinfo {year} {1991})}\BibitemShut {NoStop}%
\bibitem [{\citenamefont {Wooten}(2002)}]{wooten_structure}%
  \BibitemOpen
  \bibfield  {author} {\bibinfo {author} {\bibfnamefont {F.}~\bibnamefont
  {Wooten}},\ }\href@noop {} {\bibfield  {journal} {\bibinfo  {journal} {Acta
  Cryst. A}\ }\textbf {\bibinfo {volume} {58}},\ \bibinfo {pages} {346}
  (\bibinfo {year} {2002})}\BibitemShut {NoStop}%
\bibitem [{\citenamefont {Yuan}\ and\ \citenamefont
  {Cormack}(2002)}]{yuan_efficient}%
  \BibitemOpen
  \bibfield  {author} {\bibinfo {author} {\bibfnamefont {X.}~\bibnamefont
  {Yuan}}\ and\ \bibinfo {author} {\bibfnamefont {A.~N.}\ \bibnamefont
  {Cormack}},\ }\href@noop {} {\bibfield  {journal} {\bibinfo  {journal} {Comp.
  Mater. Sci.}\ }\textbf {\bibinfo {volume} {24}},\ \bibinfo {pages} {343}
  (\bibinfo {year} {2002})}\BibitemShut {NoStop}%
\bibitem [{\citenamefont {Roux}\ and\ \citenamefont
  {Jund}(2010)}]{leroux_ring}%
  \BibitemOpen
  \bibfield  {author} {\bibinfo {author} {\bibfnamefont {S.~L.}\ \bibnamefont
  {Roux}}\ and\ \bibinfo {author} {\bibfnamefont {P.}~\bibnamefont {Jund}},\
  }\href@noop {} {\bibfield  {journal} {\bibinfo  {journal} {Comp. Mater.
  Sci.}\ }\textbf {\bibinfo {volume} {49}},\ \bibinfo {pages} {70} (\bibinfo
  {year} {2010})}\BibitemShut {NoStop}%
\bibitem [{\citenamefont {Jin}\ \emph {et~al.}(1994)\citenamefont {Jin},
  \citenamefont {Kalia}, \citenamefont {Vashishta},\ and\ \citenamefont
  {Rino}}]{jin_structural}%
  \BibitemOpen
  \bibfield  {author} {\bibinfo {author} {\bibfnamefont {W.}~\bibnamefont
  {Jin}}, \bibinfo {author} {\bibfnamefont {R.~K.}\ \bibnamefont {Kalia}},
  \bibinfo {author} {\bibfnamefont {P.}~\bibnamefont {Vashishta}}, \ and\
  \bibinfo {author} {\bibfnamefont {J.~P.}\ \bibnamefont {Rino}},\ }\href@noop
  {} {\bibfield  {journal} {\bibinfo  {journal} {Phys. Rev. B}\ }\textbf
  {\bibinfo {volume} {118}},\ \bibinfo {pages} {50} (\bibinfo {year}
  {1994})}\BibitemShut {NoStop}%
\bibitem [{\citenamefont {Hobbs}\ \emph {et~al.}(1998)\citenamefont {Hobbs},
  \citenamefont {Jesurum}, \citenamefont {Pulim},\ and\ \citenamefont
  {Berger}}]{hobbs_local}%
  \BibitemOpen
  \bibfield  {author} {\bibinfo {author} {\bibfnamefont {L.~W.}\ \bibnamefont
  {Hobbs}}, \bibinfo {author} {\bibfnamefont {C.~E.}\ \bibnamefont {Jesurum}},
  \bibinfo {author} {\bibfnamefont {V.}~\bibnamefont {Pulim}}, \ and\ \bibinfo
  {author} {\bibfnamefont {B.}~\bibnamefont {Berger}},\ }\href@noop {}
  {\bibfield  {journal} {\bibinfo  {journal} {Philos. Mag. A}\ }\textbf
  {\bibinfo {volume} {68}},\ \bibinfo {pages} {679} (\bibinfo {year}
  {1998})}\BibitemShut {NoStop}%
\bibitem [{\citenamefont {Marians}\ and\ \citenamefont
  {Hobbs}(1988)}]{marians_characterization}%
  \BibitemOpen
  \bibfield  {author} {\bibinfo {author} {\bibfnamefont {C.~S.}\ \bibnamefont
  {Marians}}\ and\ \bibinfo {author} {\bibfnamefont {L.~W.}\ \bibnamefont
  {Hobbs}},\ }\href@noop {} {\bibfield  {journal} {\bibinfo  {journal} {J.
  Non-Cryst. Solids}\ }\textbf {\bibinfo {volume} {160}},\ \bibinfo {pages}
  {317} (\bibinfo {year} {1988})}\BibitemShut {NoStop}%
\bibitem [{\citenamefont {Marians}\ and\ \citenamefont
  {Hobbs}(1990)}]{marians_local}%
  \BibitemOpen
  \bibfield  {author} {\bibinfo {author} {\bibfnamefont {C.~S.}\ \bibnamefont
  {Marians}}\ and\ \bibinfo {author} {\bibfnamefont {L.~W.}\ \bibnamefont
  {Hobbs}},\ }\href@noop {} {\bibfield  {journal} {\bibinfo  {journal} {J.
  Non-Cryst. Solids}\ }\textbf {\bibinfo {volume} {119}},\ \bibinfo {pages}
  {269} (\bibinfo {year} {1990})}\BibitemShut {NoStop}%
\bibitem [{\citenamefont {Luzar}\ and\ \citenamefont
  {Chandler}(1996)}]{chandler_HB}%
  \BibitemOpen
  \bibfield  {author} {\bibinfo {author} {\bibfnamefont {A.}~\bibnamefont
  {Luzar}}\ and\ \bibinfo {author} {\bibfnamefont {D.}~\bibnamefont
  {Chandler}},\ }\href@noop {} {\bibfield  {journal} {\bibinfo  {journal}
  {Nature}\ }\textbf {\bibinfo {volume} {379}},\ \bibinfo {pages} {55}
  (\bibinfo {year} {1996})}\BibitemShut {NoStop}%
\bibitem [{\citenamefont {Prada-Gracia}, \citenamefont {Shevchuk},\ and\
  \citenamefont {Rao}(2013)}]{prada_2013}%
  \BibitemOpen
  \bibfield  {author} {\bibinfo {author} {\bibfnamefont {D.}~\bibnamefont
  {Prada-Gracia}}, \bibinfo {author} {\bibfnamefont {R.}~\bibnamefont
  {Shevchuk}}, \ and\ \bibinfo {author} {\bibfnamefont {F.}~\bibnamefont
  {Rao}},\ }\href@noop {} {\bibfield  {journal} {\bibinfo  {journal} {J. Chem.
  Phys.}\ }\textbf {\bibinfo {volume} {139}},\ \bibinfo {pages} {084501}
  (\bibinfo {year} {2013})}\BibitemShut {NoStop}%
\bibitem [{\citenamefont {Shi}, \citenamefont {Russo},\ and\ \citenamefont
  {Tanaka}(2018{\natexlab{b}})}]{shi_2018_2}%
  \BibitemOpen
  \bibfield  {author} {\bibinfo {author} {\bibfnamefont {R.}~\bibnamefont
  {Shi}}, \bibinfo {author} {\bibfnamefont {J.}~\bibnamefont {Russo}}, \ and\
  \bibinfo {author} {\bibfnamefont {H.}~\bibnamefont {Tanaka}},\ }\href@noop {}
  {\bibfield  {journal} {\bibinfo  {journal} {J. Chem. Phys.}\ }\textbf
  {\bibinfo {volume} {149}},\ \bibinfo {pages} {224502} (\bibinfo {year}
  {2018}{\natexlab{b}})}\BibitemShut {NoStop}%
\bibitem [{\citenamefont {Li}\ \emph {et~al.}(2020{\natexlab{b}})\citenamefont
  {Li}, \citenamefont {Zhong}, \citenamefont {Yan}, \citenamefont {Zhang},
  \citenamefont {Xu}, \citenamefont {Francisco},\ and\ \citenamefont
  {Zeng}}]{LiClathrate2020}%
  \BibitemOpen
  \bibfield  {author} {\bibinfo {author} {\bibfnamefont {L.}~\bibnamefont
  {Li}}, \bibinfo {author} {\bibfnamefont {J.}~\bibnamefont {Zhong}}, \bibinfo
  {author} {\bibfnamefont {Y.}~\bibnamefont {Yan}}, \bibinfo {author}
  {\bibfnamefont {J.}~\bibnamefont {Zhang}}, \bibinfo {author} {\bibfnamefont
  {J.}~\bibnamefont {Xu}}, \bibinfo {author} {\bibfnamefont {J.~S.}\
  \bibnamefont {Francisco}}, \ and\ \bibinfo {author} {\bibfnamefont {X.~C.}\
  \bibnamefont {Zeng}},\ }\href@noop {} {\bibfield  {journal} {\bibinfo
  {journal} {Proc. Natl. Acad. Sci. USA}\ }\textbf {\bibinfo {volume} {117}},\
  \bibinfo {pages} {24701} (\bibinfo {year} {2020}{\natexlab{b}})}\BibitemShut
  {NoStop}%
\bibitem [{\citenamefont {de~Oca}\ \emph {et~al.}(2020)\citenamefont {de~Oca},
  \citenamefont {Sciortino}, ,\ and\ \citenamefont
  {Appignanesi}}]{MontesStructural2020}%
  \BibitemOpen
  \bibfield  {author} {\bibinfo {author} {\bibfnamefont {J.~M.~M.}\
  \bibnamefont {de~Oca}}, \bibinfo {author} {\bibfnamefont {F.}~\bibnamefont
  {Sciortino}}, , \ and\ \bibinfo {author} {\bibfnamefont {G.~A.}\ \bibnamefont
  {Appignanesi}},\ }\href@noop {} {\bibfield  {journal} {\bibinfo  {journal}
  {J. Chem. Phys.}\ }\textbf {\bibinfo {volume} {152}},\ \bibinfo {pages}
  {244503} (\bibinfo {year} {2020})}\BibitemShut {NoStop}%
\bibitem [{\citenamefont {Skinner}\ \emph {et~al.}(2013)\citenamefont
  {Skinner}, \citenamefont {Huang}, \citenamefont {Schlesinger}, \citenamefont
  {Pettersson}, \citenamefont {Nilsson},\ and\ \citenamefont
  {Benmore}}]{skinnerExp}%
  \BibitemOpen
  \bibfield  {author} {\bibinfo {author} {\bibfnamefont {L.~B.}\ \bibnamefont
  {Skinner}}, \bibinfo {author} {\bibfnamefont {C.}~\bibnamefont {Huang}},
  \bibinfo {author} {\bibfnamefont {D.}~\bibnamefont {Schlesinger}}, \bibinfo
  {author} {\bibfnamefont {L.~G.~M.}\ \bibnamefont {Pettersson}}, \bibinfo
  {author} {\bibfnamefont {A.}~\bibnamefont {Nilsson}}, \ and\ \bibinfo
  {author} {\bibfnamefont {C.~J.}\ \bibnamefont {Benmore}},\ }\href@noop {}
  {\bibfield  {journal} {\bibinfo  {journal} {J. Chem. Phys.}\ }\textbf
  {\bibinfo {volume} {138}},\ \bibinfo {pages} {074506} (\bibinfo {year}
  {2013})}\BibitemShut {NoStop}%
\bibitem [{\citenamefont {Soper}\ and\ \citenamefont
  {Benmore}(2008)}]{soperExp}%
  \BibitemOpen
  \bibfield  {author} {\bibinfo {author} {\bibfnamefont {A.~K.}\ \bibnamefont
  {Soper}}\ and\ \bibinfo {author} {\bibfnamefont {C.~J.}\ \bibnamefont
  {Benmore}},\ }\href@noop {} {\bibfield  {journal} {\bibinfo  {journal} {Phys.
  Rev. Lett.}\ }\textbf {\bibinfo {volume} {101}},\ \bibinfo {pages} {065502}
  (\bibinfo {year} {2008})}\BibitemShut {NoStop}%
\bibitem [{\citenamefont {Praprotnik}, \citenamefont {Jane\v{z}i\v{c}},\ and\
  \citenamefont {Mavri}(2004)}]{praprotnik2004}%
  \BibitemOpen
  \bibfield  {author} {\bibinfo {author} {\bibfnamefont {M.}~\bibnamefont
  {Praprotnik}}, \bibinfo {author} {\bibfnamefont {D.}~\bibnamefont
  {Jane\v{z}i\v{c}}}, \ and\ \bibinfo {author} {\bibfnamefont {J.}~\bibnamefont
  {Mavri}},\ }\href@noop {} {\bibfield  {journal} {\bibinfo  {journal} {J.
  Phys. Chem. A}\ }\textbf {\bibinfo {volume} {108}},\ \bibinfo {pages} {11056}
  (\bibinfo {year} {2004})}\BibitemShut {NoStop}%
\bibitem [{\citenamefont {{MacKerell Jr.}}\ \emph {et~al.}(1998)\citenamefont
  {{MacKerell Jr.}}, \citenamefont {Bashford}, \citenamefont {Bellott},
  \citenamefont {Jr}, \citenamefont {Evanseck}, \citenamefont {Field},
  \citenamefont {Fischer}, \citenamefont {Gao}, \citenamefont {Guo},
  \citenamefont {Ha}, \citenamefont {Joseph-McCarthy}, \citenamefont {Kuchnir},
  \citenamefont {Kuczera}, \citenamefont {Lau}, \citenamefont {Mattos},
  \citenamefont {Michnick}, \citenamefont {Ngo}, \citenamefont {Nguyen},
  \citenamefont {Prodhom}, \citenamefont {Reiher}, \citenamefont {Roux},
  \citenamefont {Schlenkrich}, \citenamefont {Smith}, \citenamefont {Stote},
  \citenamefont {Straub}, \citenamefont {Watanabe}, \citenamefont
  {Wi\'{o}rkiewicz-Kuczera}, \citenamefont {Yin},\ and\ \citenamefont
  {Karplus}}]{tip3pcharmm}%
  \BibitemOpen
  \bibfield  {author} {\bibinfo {author} {\bibfnamefont {A.~D.}\ \bibnamefont
  {{MacKerell Jr.}}}, \bibinfo {author} {\bibfnamefont {D.}~\bibnamefont
  {Bashford}}, \bibinfo {author} {\bibfnamefont {M.}~\bibnamefont {Bellott}},
  \bibinfo {author} {\bibfnamefont {R.~L.~D.}\ \bibnamefont {Jr}}, \bibinfo
  {author} {\bibfnamefont {J.~D.}\ \bibnamefont {Evanseck}}, \bibinfo {author}
  {\bibfnamefont {M.~J.}\ \bibnamefont {Field}}, \bibinfo {author}
  {\bibfnamefont {S.}~\bibnamefont {Fischer}}, \bibinfo {author} {\bibfnamefont
  {J.}~\bibnamefont {Gao}}, \bibinfo {author} {\bibfnamefont {H.}~\bibnamefont
  {Guo}}, \bibinfo {author} {\bibfnamefont {S.}~\bibnamefont {Ha}}, \bibinfo
  {author} {\bibfnamefont {D.}~\bibnamefont {Joseph-McCarthy}}, \bibinfo
  {author} {\bibfnamefont {L.}~\bibnamefont {Kuchnir}}, \bibinfo {author}
  {\bibfnamefont {K.}~\bibnamefont {Kuczera}}, \bibinfo {author} {\bibfnamefont
  {F.~T.~K.}\ \bibnamefont {Lau}}, \bibinfo {author} {\bibfnamefont
  {C.}~\bibnamefont {Mattos}}, \bibinfo {author} {\bibfnamefont
  {S.}~\bibnamefont {Michnick}}, \bibinfo {author} {\bibfnamefont
  {T.}~\bibnamefont {Ngo}}, \bibinfo {author} {\bibfnamefont {D.~T.}\
  \bibnamefont {Nguyen}}, \bibinfo {author} {\bibfnamefont {B.}~\bibnamefont
  {Prodhom}}, \bibinfo {author} {\bibfnamefont {W.~E.}\ \bibnamefont {Reiher}},
  \bibinfo {author} {\bibfnamefont {B.}~\bibnamefont {Roux}}, \bibinfo {author}
  {\bibfnamefont {M.}~\bibnamefont {Schlenkrich}}, \bibinfo {author}
  {\bibfnamefont {J.~C.}\ \bibnamefont {Smith}}, \bibinfo {author}
  {\bibfnamefont {R.}~\bibnamefont {Stote}}, \bibinfo {author} {\bibfnamefont
  {J.}~\bibnamefont {Straub}}, \bibinfo {author} {\bibfnamefont
  {M.}~\bibnamefont {Watanabe}}, \bibinfo {author} {\bibfnamefont
  {J.}~\bibnamefont {Wi\'{o}rkiewicz-Kuczera}}, \bibinfo {author}
  {\bibfnamefont {D.}~\bibnamefont {Yin}}, \ and\ \bibinfo {author}
  {\bibfnamefont {M.}~\bibnamefont {Karplus}},\ }\href@noop {} {\bibfield
  {journal} {\bibinfo  {journal} {J. Phys. Chem. B}\ }\textbf {\bibinfo
  {volume} {102}},\ \bibinfo {pages} {3586} (\bibinfo {year}
  {1998})}\BibitemShut {NoStop}%
\bibitem [{\citenamefont {Calero}, \citenamefont {Stanley},\ and\ \citenamefont
  {Franzese}(2016)}]{CaleroStructural2016}%
  \BibitemOpen
  \bibfield  {author} {\bibinfo {author} {\bibfnamefont {C.}~\bibnamefont
  {Calero}}, \bibinfo {author} {\bibfnamefont {E.~H.}\ \bibnamefont {Stanley}},
  \ and\ \bibinfo {author} {\bibfnamefont {G.}~\bibnamefont {Franzese}},\
  }\href@noop {} {\bibfield  {journal} {\bibinfo  {journal} {Materials}\
  }\textbf {\bibinfo {volume} {9}},\ \bibinfo {pages} {319} (\bibinfo {year}
  {2016})}\BibitemShut {NoStop}%
\bibitem [{\citenamefont {Samatas}\ \emph {et~al.}(2018)\citenamefont
  {Samatas}, \citenamefont {Calero}, \citenamefont {Martelli},\ and\
  \citenamefont {Franzese}}]{samatas_2018}%
  \BibitemOpen
  \bibfield  {author} {\bibinfo {author} {\bibfnamefont {S.}~\bibnamefont
  {Samatas}}, \bibinfo {author} {\bibfnamefont {C.}~\bibnamefont {Calero}},
  \bibinfo {author} {\bibfnamefont {F.}~\bibnamefont {Martelli}}, \ and\
  \bibinfo {author} {\bibfnamefont {G.}~\bibnamefont {Franzese}},\ }\href@noop
  {} {\bibfield  {journal} {\bibinfo  {journal} {arXiv:1811.01911
  [cond-mat.soft]}\ } (\bibinfo {year} {2018})}\BibitemShut {NoStop}%
\bibitem [{\citenamefont {Gallo}, \citenamefont {Rovere},\ and\ \citenamefont
  {Chen}(2010)}]{gallo_2010}%
  \BibitemOpen
  \bibfield  {author} {\bibinfo {author} {\bibfnamefont {P.}~\bibnamefont
  {Gallo}}, \bibinfo {author} {\bibfnamefont {M.}~\bibnamefont {Rovere}}, \
  and\ \bibinfo {author} {\bibfnamefont {S.-H.}\ \bibnamefont {Chen}},\
  }\href@noop {} {\bibfield  {journal} {\bibinfo  {journal} {J. Phys. Chem.
  Lett.}\ }\textbf {\bibinfo {volume} {1}},\ \bibinfo {pages} {729} (\bibinfo
  {year} {2010})}\BibitemShut {NoStop}%
\bibitem [{\citenamefont {Camisasca}, \citenamefont {Marzio},\ and\
  \citenamefont {Gallo}(2020)}]{CamisascaEffect2020}%
  \BibitemOpen
  \bibfield  {author} {\bibinfo {author} {\bibfnamefont {G.}~\bibnamefont
  {Camisasca}}, \bibinfo {author} {\bibfnamefont {M.~D.}\ \bibnamefont
  {Marzio}}, \ and\ \bibinfo {author} {\bibfnamefont {P.}~\bibnamefont
  {Gallo}},\ }\href@noop {} {\bibfield  {journal} {\bibinfo  {journal} {J.
  Chem. Phys.}\ }\textbf {\bibinfo {volume} {153}},\ \bibinfo {pages} {224503}
  (\bibinfo {year} {2020})}\BibitemShut {NoStop}%
\bibitem [{\citenamefont {Iorio}, \citenamefont {Camisasca},\ and\
  \citenamefont {Gallo}(2019)}]{IorioGlassy2019}%
  \BibitemOpen
  \bibfield  {author} {\bibinfo {author} {\bibfnamefont {A.}~\bibnamefont
  {Iorio}}, \bibinfo {author} {\bibfnamefont {G.}~\bibnamefont {Camisasca}}, \
  and\ \bibinfo {author} {\bibfnamefont {P.}~\bibnamefont {Gallo}},\
  }\href@noop {} {\bibfield  {journal} {\bibinfo  {journal} {Sci. China Phys.
  Mech. Astron.}\ }\textbf {\bibinfo {volume} {62}},\ \bibinfo {pages} {107011}
  (\bibinfo {year} {2019})}\BibitemShut {NoStop}%
\bibitem [{\citenamefont {Iorio}\ \emph {et~al.}(2020)\citenamefont {Iorio},
  \citenamefont {Minozzi}, \citenamefont {Camisasca}, \citenamefont {Rovere},\
  and\ \citenamefont {Gallo}}]{IorioSlow2020}%
  \BibitemOpen
  \bibfield  {author} {\bibinfo {author} {\bibfnamefont {A.}~\bibnamefont
  {Iorio}}, \bibinfo {author} {\bibfnamefont {M.}~\bibnamefont {Minozzi}},
  \bibinfo {author} {\bibfnamefont {G.}~\bibnamefont {Camisasca}}, \bibinfo
  {author} {\bibfnamefont {M.}~\bibnamefont {Rovere}}, \ and\ \bibinfo {author}
  {\bibfnamefont {P.}~\bibnamefont {Gallo}},\ }\href@noop {} {\bibfield
  {journal} {\bibinfo  {journal} {Philosophical Magazine}\ }\textbf {\bibinfo
  {volume} {100}},\ \bibinfo {pages} {2582} (\bibinfo {year}
  {2020})}\BibitemShut {NoStop}%
\bibitem [{\citenamefont {Tenuzzo}, \citenamefont {Camisasca},\ and\
  \citenamefont {Gallo}(2020)}]{TenuzzoProtein2020}%
  \BibitemOpen
  \bibfield  {author} {\bibinfo {author} {\bibfnamefont {L.}~\bibnamefont
  {Tenuzzo}}, \bibinfo {author} {\bibfnamefont {G.}~\bibnamefont {Camisasca}},
  \ and\ \bibinfo {author} {\bibfnamefont {P.}~\bibnamefont {Gallo}},\
  }\href@noop {} {\bibfield  {journal} {\bibinfo  {journal} {Molecules}\
  }\textbf {\bibinfo {volume} {25}},\ \bibinfo {pages} {4570} (\bibinfo {year}
  {2020})}\BibitemShut {NoStop}%
\bibitem [{\citenamefont {Calero}\ and\ \citenamefont
  {Franzese}(2020)}]{CaleroWater2020}%
  \BibitemOpen
  \bibfield  {author} {\bibinfo {author} {\bibfnamefont {C.}~\bibnamefont
  {Calero}}\ and\ \bibinfo {author} {\bibfnamefont {G.}~\bibnamefont
  {Franzese}},\ }\href@noop {} {\bibfield  {journal} {\bibinfo  {journal} {J.
  Mol. Liq.}\ }\textbf {\bibinfo {volume} {317}},\ \bibinfo {pages} {114027}
  (\bibinfo {year} {2020})}\BibitemShut {NoStop}%
\end{thebibliography}%

\end{document}